\documentclass
[twocolumn,aps,preprintnumbers,prx,superscriptaddress,showpacs]{revtex4}%
\usepackage{graphicx}
\usepackage{dcolumn}
\usepackage{bm}
\usepackage{times}
\usepackage[colorlinks=true,linkcolor=blue,anchorcolor=blue,citecolor=blue,urlcolor=blue]%
{hyperref}
\usepackage{multirow}
\usepackage{amsmath}
\usepackage{amsfonts}
\usepackage{amssymb}%
\begin{document}
\title{Topological superconducting states in monolayer FeSe/SrTiO$_{3}$}
\author{Ningning Hao}
\affiliation{Department of Physics, The University of Hong Kong, Pokfulam Road, Hong Kong, China}
\author{Shun-Qing Shen}
\affiliation{Department of Physics, The University of Hong Kong, Pokfulam Road, Hong Kong, China}

\begin{abstract}
The monolayer FeSe with a thickness of one unit cell grown on a single-crystal
SrTiO$_{3}$ substrate (FeSe/STO) exhibits striking high-temperature
superconductivity with transition temperature $T_{c}$ over 65K reported by
recent experimental measurements. In this work, through analyzing the
distinctive electronic structure, and providing systematic classification of
the pairing symmetry , we find that both $s$-and $p$-wave pairing with odd
parity give rise to topological superconducting states in monolayer FeSe, and
the exotic properties of $s$-wave topological superconducting states have
close relations with the unique non-symmorphic lattice structure which induces
the orbital-momentum locking. Our results indicate that the monolayer FeSe
could be in the topological nontrivial $s$-wave superconducting states if the
relevant effective pairing interactions are dominant in comparison with other candidates.

\end{abstract}

\maketitle

%\email{sshen@hku.hk}

%insert suggested PACS numbers in braces on next line%\maketitle must follow title, authors, abstract, \pacs, and \keywords

\section{Introduction}

Topological superconductors\cite{Qi2011RMP,Fu2008PRL,Sau2010PRL,Fu2010PRL} and
iron-based superconductors\cite{Yoichi2008JACS} have been research focuses of
condensed matter physics in recent years. Topological superconductors have a
full pairing gap in the bulk and gapless surface or edge Andreev bound states
known as Majorana fermions. Recent scanning tunneling microscopy/spectroscopy
(STM/S) measurements observed a robust zero-energy bound state at randomly
distributed interstitial excess Fe sites in superconducting Fe(Te,Se), and the
behavior of zero-energy bound state resembles the Majorana
fermion\cite{Yin2015NP}. Theoretically, one possible scenario accounting for
this puzzle is that Fe(Te,Se) could be in a topological superconducting (SC)
state. If it is the case, we can expect that nontrivial topology can integrate
into the SC states in iron-based superconductors.

Recently, some studies\cite{Hao2014PRX,Wu2014arXiv} have revealed that the
band structures can be tuned to have nontrivial topological properties in
monolayer Fe(Te,Se) and monolayer FeSe/STO. Furthermore, in electron-doped
monolayer FeSe/STO, the experimental measurements have observed high
temperature superconductivity with $T_{c}$ over
65K\cite{Wang2012CPL,Liu2012NC,He2013NM,Tan2013NM,Peng2014NC,Zhang2014CPL,Lee2014N,Ge2015NM}%
. In analogy to the doped topological insulators, which are strongly believed
to be topological
superconductors\cite{Fu2010PRL,Hor2010PRL,Sasaki2011PRL,Sasaki2012PRL}, a
natural question arises, can the electron-doped monolayer FeSe/STO be
topological superconductors?

In this paper, we propose that the electron-doped monolayer FeSe/STO could be
an odd-parity topological superconductor in the spin-triplet orbital-singlet
$s$-wave pairing channel. To show this exotic state, we first analyse the
distinctive electronic structure of monolayer FeSe/STO, and present a
systematic classification of the pairing symmetry in monolayer FeSe/STO from
the lattice symmetric group. Second, we discuss the topological properties of
such odd-parity SC states, and extract the minimum effective models to capture
the essential physics. Third, we calculate the phase diagram of SC states
according to different scenarios of effective pairing interaction. Finally, we
discuss the experimental signatures of the topological SC states.

\begin{figure}[ptb]
\begin{center}
\includegraphics[width=1\linewidth]{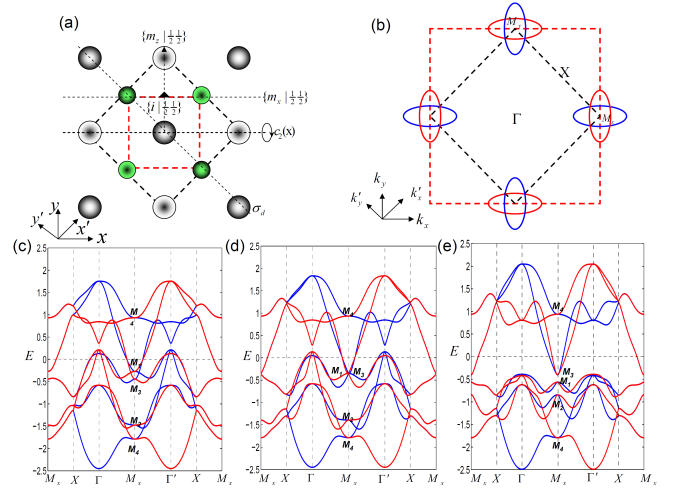}
\end{center}
\caption{ (a) The Se-Fe-Se trilayer structure. The black/green balls with deep
and light filling label Fe/Se atoms. Here, the deep/light filling Se atoms are
above/below the Fe plane. The red/black dashed squares label the one-Fe/two-Fe
unit cells. (b) The Fermi surface of monolayer FeSe/STO is schematically
illustrated. The red/blue electron pockets have odd/even orbital parity. The
red/black dashed squares label the one-Fe/two-Fe Brillouin zone.The evolution
of the band structure from (c) the free-standing monolayer FeSe to (e)
monolayer FeSe/STO with tensile strain. The red/blue color labels the spectrum
with odd/even orbital parity. }%
\end{figure}

\section{Pairing symmetry classifications}

The lattice structure of monolayer FeSe is shown in Fig. 1(a). The two-Fe unit
cell includes two Se and two Fe labeled by A and B. The space group $P4/nmm$
governs the Se-Fe-Se trilayer structure, and belongs to a non-symmorphic
group\cite{Hu2013PRX,Hao2014PRB,Subedi2008PRB,Cvetkovic}. Indeed, there exists
a $n$-glide plane described by the operator $\{m_{z}|\frac{1}{2}\frac{1}{2}%
\}$, which involves a fractional translation $(\frac{1}{2}\frac{1}{2})$
combining with the $ab$-plane mirror. Centered on an Fe atom (see Fig. 1(a)),
eight point group operations, $E$, $2S_{4}$, $c_{2}(z)$, $c_{2}(x)$,
$c_{2}(y)$ and $2\sigma_{d}$, form a $D_{2d}$ point group. Together with an
inversion followed by fractional translations $(\frac{1}{2}\frac{1}{2})$,
i.e., $\{i|\frac{1}{2}\frac{1}{2}\}$, they generate all the elements of
$P4/nmm$. The 16 operations do not form a point group. However, if the
fractional translation $(\frac{1}{2}\frac{1}{2})$ is stripped off, the 16
operations form a point group, which indeed is $D_{4h}$. It is convenient to
classify the pairing symmetry with the irreducible representation (IR) of
$D_{4h}$. For this purpose, one simple way is to recompose the Bloch wave
functions in the one-Fe Brillouin zone (BZ).

The glide plane symmetry $\{m_{z}|\frac{1}{2}\frac{1}{2}\}$ divides the five
$d$ orbitals into two groups ($d_{xz},d_{yz}$) and ($d_{xy},d_{x^{2}-y^{2}%
},d_{z^{2}}$), and each group is recomposed to be the eigen-states of the
glide plane operation with the definite orbital parities. The tight-binding
Hamiltonian can also be decomposed into two parts with inverse orbital
parities, which allow us to transfer the two-Fe unit cell picture into one-Fe
unit cell picture\cite{Hu2013PRX,Hao2014PRB,Subedi2008PRB}. In momentum space,
the tight-binding Hamiltonian in one-Fe unit cell picture can be written as
\begin{equation}
H_{0}=\sum_{\mathbf{k},\sigma}\psi_{\sigma}^{o\dag}(\mathbf{k})A_{o}%
(k)\psi_{\sigma}^{o}(\mathbf{k})+\sum_{\mathbf{k},\sigma}\psi_{\sigma}^{e\dag
}(\mathbf{k})A_{e}(k)\psi_{\sigma}^{e}(\mathbf{k}). \label{TB1}%
\end{equation}
Here, the first/second term has odd/even orbital parity under the glide plane
operation. $\psi_{\sigma}^{o}(k)=[d_{xz,\sigma}(\mathbf{k})$, $d_{yz,\sigma
}(\mathbf{k})$, $d_{x^{2}-y^{2},\sigma}(\mathbf{k})$, $d_{xy,\sigma
}(\mathbf{k})$, $d_{z^{2},\sigma}(\mathbf{k})]^{T}$ with $d_{m,\sigma
}(\mathbf{k})$ denoting the electron annihilation operator at the $m$th
orbital with momentum $\mathbf{k}$ and spin $\sigma$. $\psi_{\sigma}%
^{e}(\mathbf{k})=\psi_{\sigma}^{o}(\mathbf{k}+\mathbf{Q})$ and $A_{e}%
(\mathbf{k})=A_{o}(\mathbf{k}+\mathbf{Q})$ with $\mathbf{Q}=(\pi,\pi)$ (see
Appendix A for details). The energy spectra from Eq.(\ref{TB1}) are shown in
Fig. 1, in which Fig. 1(e) is consistent with observations of the
angle-resolved photoemission spectroscopy (ARPES)\cite{Liu2012NC,He2013NM},
and the chemical potential is set to satisfy that 10\% electrons is doped per
Fe clarified by experiments\cite{Liu2012NC,He2013NM,Tan2013NM}. The
fundamental difference between Fig. 1(c) and (f) is referred to the
band-renormalization effect induced by the strain from the STO substrate,
which strongly modulates the hopping parameters between the ($d_{xz}%
,d_{yz},d_{xy}$) orbitals and switches the positions of two doubly degenerate
points $M_{1}$ and $M_{3}$ at $M_{x}$ high symmetric point, where the $M_{1}$
point mainly has ($d_{xz},d_{yz}$) orbital weight and the $M_{3}$ point mainly
has $d_{xy}$ orbital weight. This picture is the most natural and simplest to
account for the distinctive electronic structure of monolayer FeSe/STO
compared to other scenarios\cite{Liu2012PRB,Timur2013JPCM,Zheng2013SR}.

The SC order parameters should follow the IRs of the symmetry group of the
system. It is safe to use $D_{4h}$ to do so in the picture of one-Fe unit cell
according to our aforementioned arguments. There exist two kinds of
symmetry-allowed Cooper pairs, i.e., $(\mathbf{k},-\mathbf{k})$ and
$(\mathbf{k},-\mathbf{k}+\mathbf{Q})$ pairing channels. Previously, the
$(\mathbf{k},-\mathbf{k}+\mathbf{Q})$ pairing channels are proposed to coexist
with $(\mathbf{k},-\mathbf{k})$ pairing channels to explain the nodeless and
sign-change gap structures in iron-based
superconductors\cite{Hu2013PRX,Hao2014PRB}. The price for coexistence of both
kinds of pairings is that the orbital parities are mixed and the spatial
inversion symmetry is broken. Here we focus on an SC state with only one IRs
in $(\mathbf{k},-\mathbf{k})$ pairing channel and leave to discuss the
irrelevant $(\mathbf{k},-\mathbf{k}+\mathbf{Q})$ pairing channel in Appendix
B. Moreover, we only need to consider the pairings between the three $t_{2g}$
orbitals as the orbital weight for $E_{g}$ orbitals are neglectable on the
Fermi surfaces\cite{Cao2008PRB}. Define the Nambu basis, $\Psi(\mathbf{k}%
)=[\{d_{\uparrow}(\mathbf{k})\},\{d_{\downarrow}(\mathbf{k})\},\{d_{\downarrow
}^{\dag}(-\mathbf{k})\},\{-d_{\uparrow}^{\dag}(-\mathbf{k})\}]^{T}$ with
$\{d_{\sigma}(\mathbf{k})\}=\{d_{xz,\sigma}(\mathbf{k}),d_{yz,\sigma
}(\mathbf{k}),d_{xy,\sigma}(\mathbf{k})\}$. The pairing term in the
Bogoliubov--de Gennes (BdG) Hamiltonian can be expressed as
\begin{equation}
H_{p}=\sum_{\mathbf{k}}\Psi^{\dag}(\mathbf{k})\Delta(\mathbf{k})\tau_{x}%
\Psi(\mathbf{k}). \label{Hp}%
\end{equation}
Here, $\tau_{x}$ is one Pauli matrix in Nambu space, and $\Delta(\mathbf{k})$
is a $6\times6$ matrix. Our purpose is to identify the exact form of
$\Delta(\mathbf{k})$. For convenience, we utilize four Pauli matrices
($s_{0},s_{x},s_{y},s_{z}$) to span spin space and nine Gell-Mann matrices
($\lambda_{0},...,\lambda_{8}$) (see Appendix B for definitions of Gell-Mann
matrices) to span orbital space. In such a way, $\Delta(\mathbf{k})$ can be
decomposed into the product of the Pauli matrices and Gell-Mann matrices, i.e.
$\Delta(\mathbf{k})=f(\mathbf{k})s_{m}\lambda_{n}$ in which $f(\mathbf{k})$ is
the pairing form factor. We summarize all the possibilities of the
$(\mathbf{k},-\mathbf{k})$ on-site pairing channels according to the IRs of
$D_{4h}$ in Table \ref{pairk-k} and non-on-site pairing channels up to the
next-nearest neighbor in Table \ref{pair-long}. In both Tables \ref{pairk-k}
and \ref{pair-long}, the spin-singlet/-triplet pairing channels are listed in
the first/second parts.

\begin{center}
\begin{table}[ptb]
\caption{The IRs of all the possible onsite superconducting pairing in
$(\mathbf{k},\mathbf{-k})$ channels. Here, $\eta_{1/4}=\mp\frac{1}{3}%
(\lambda_{0}+2\sqrt{3}\lambda_{8})$ and $\eta_{2/3}=\frac{1}{3}(\mp\lambda
_{0}\pm\sqrt{3}\lambda_{8}\mp3\lambda_{3/1})$. }%
\label{pairk-k}%
\begin{tabular}
[c]{llllll}\hline\hline
$(\mathbf{k},-\mathbf{k}):\Delta(\mathbf{k})$ & $c_{2}(z)$ & $c_{2}(x)$ &
$\ \ \ \sigma_{d}$ & $\{i|\frac{1}{2}\frac{1}{2}\}$ & $IR$\\\hline
& $-is_{z}\eta_{1}$ & $-is_{x}\eta_{2}$ & $\frac{-i(s_{x}-s_{y})\eta_{3}%
}{\sqrt{2}}$ & $s_{0}\eta_{4}$ & \\
$s_{0}\lambda_{0}$ & 1 & 1 & 1 & 1 & $A_{1g}^{(1)}$\\
$s_{0}\lambda_{8}$ & 1 & 1 & 1 & 1 & $A_{1g}$\\
$s_{0}\lambda_{1}$ & 1 & -1 & 1 & 1 & $B_{2g}$\\
$s_{0}(\lambda_{4},\lambda_{6})$ & (-1,-1) & (1,-1) & $s_{0}(\lambda
_{6},\lambda_{4})$ & (-1,-1) & $E_{u}$\\\hline
$is_{z}\lambda_{2}$ & 1 & 1 & 1 & 1 & $A_{1g}$\\
$s_{z}(\lambda_{5},\lambda_{7})$ & (-1,-1) & (-1,1) & -$s_{z}(\lambda
_{7},\lambda_{5})$ & (-1,-1) & $E_{u}^{(1)}$\\
$i(s_{x},s_{y})\lambda_{2}$ & (-1,-1) & (-1,1) & $i(s_{y},s_{x})\lambda_{2}$ &
(1,1) & $E_{g}$\\
$i(s_{x}\lambda_{5},s_{y}\lambda_{7})$ & (1,1) & (1,1) & -$i(s_{y}\lambda
_{7},s_{x}\lambda_{5})$ & (-1,-1) & $E_{u}^{(2)}$\\
$i(s_{y}\lambda_{5},s_{x}\lambda_{7})$ & (1,1) & (-1,-1) & -$i(s_{x}%
\lambda_{7},s_{y}\lambda_{5})$ & (-1,-1) & $E_{u}^{(2^{\prime})}%
$\\\hline\hline
\end{tabular}
\end{table}
\end{center}

\begin{table}[ptb]
\caption{The IRs of all the possible nearest and next nearest neighbor
superconducting pairing in $\mathbf{(k},\mathbf{-k)}$ channels. Here,
$f_{1/2}(k)=\cos k_{x}\pm\cos k_{y}$; $f_{4}(k)=\cos k_{x}\cos k_{y}$;
$[f_{3}(k_{x}),f_{3}(k_{y})]=[\sin k_{x},\sin k_{y}]$; $f_{5}(k)=\sin
k_{x}\sin k_{y}$.}%
\label{pair-long}%
\begin{tabular}
[c]{ll}\hline\hline
$(\mathbf{k},-\mathbf{k}):\Delta(\mathbf{k})$ & IR\\\hline
$f_{1/4}(k)s_{0}\lambda_{0/8},f_{5}(k)s_{0}\lambda_{1},f_{3}(k_{x}%
)s_{0}\lambda_{5}+f_{3}(k_{y})s_{0}\lambda_{7}$ & $A_{1g}^{(2)}$\\
$f_{2}(k)s_{0}\lambda_{0/8},f_{3}(k_{x})s_{0}\lambda_{5}-f_{3}(k_{y}%
)s_{0}\lambda_{7}$ & $B_{1g}^{(1)}$\\
$f_{2}(k)s_{0}\lambda_{1},f_{3}(k_{y})s_{0}\lambda_{5}-f_{3}(k_{x}%
)s_{0}\lambda_{7}$ & $A_{2g}$\\
$f_{5}(k)s_{0}\lambda_{0/8},f_{1/4}(k)s_{0}\lambda_{1},f_{3}(k_{y}%
)s_{0}\lambda_{5}+f_{3}(k_{x})s_{0}\lambda_{7}$ & $B_{2g}$\\\hline
$if_{1/4}(k)s_{z}\lambda_{2},i^{1/0/0}[f_{3}(k_{x})s_{z/x/y}\lambda_{4}%
+if_{3}(k_{y})s_{z/y/x}\lambda_{6}]$ & $A_{1g}$\\
$if_{2}(k)s_{z}\lambda_{2},i^{1/0/0}[f_{3}(k_{x})s_{z/x/y}\lambda_{4}%
-if_{3}(k_{y})s_{z/y/x}\lambda_{6}]$ & $B_{1g}$\\
$i^{1/0/0}[f_{3}(k_{y})s_{z/x/y}\lambda_{4}-f_{3}(k_{x})s_{z/y/x}\lambda_{6}]$
& $A_{2g}$\\
$if_{5}(k)s_{z}\lambda_{2},i^{1/0/0}[f_{3}(k_{y})s_{z/x/y}\lambda_{4}%
+f_{3}(k_{x})s_{z/y/x}\lambda_{6}]$ & $B_{2g}$\\
$if_{1/2/4/5}(k)(s_{x},s_{y})\lambda_{2}$ & $E_{g}$\\
$f_{3}(k_{x})s_{x/y}\lambda_{0}\pm f_{3}(k_{y})s_{y/x}\lambda_{0}$ &
$A_{1u}^{(1)}$\\
$\lbrack f_{3}(k_{x}),f_{3}(k_{y})]s_{z}\lambda_{0}$ & $E_{u}^{(3)}%
$\\\hline\hline
\end{tabular}
\end{table}

\section{Topological superconducting states}

To evaluate the pairing channels that could support the topological SC states,
we first impose the nodeless gap structures restrictions to the pairing
channels in Tables \ref{pairk-k} and \ref{pair-long} according to ARPES and
STM/S experimental results\cite{Wang2012CPL,Liu2012NC,He2013NM}, i.e.,
$A_{1g}^{(1)}$, $E_{u}^{(1)}$, $E_{u}^{(2)}$ and $E_{u}^{(2^{\prime})}$ in
Table \ref{pairk-k} and $A_{1g}^{(1)}$ with $f_{4}(k)s_{0}\lambda_{0}$,
$B_{1g}^{(1)}$, $A_{1u}^{(1)}$ and $E_{u}^{(3)}$ in Table \ref{pair-long}.
Second, we focus on the odd-parity pairing channels based on the proposals
that odd parity pairings usually support the topological SC states in doped
topological insulators\cite{Fu2010PRL}. Finally, we consider the SC states
with the $C_{4}$ rotation symmetry verified by both experimental
observations\cite{Liu2012NC,He2013NM,Tan2013NM,Peng2014NC} and our
calculations in Section IV. This constraint forces the time-reversal (TR)
symmetry to be broken spontaneously for some $E_{u}$ states. With all above
constraints and turn to the monolayer FeSe/STO, four possible odd-parity
pairing states survive: (1) $E_{u}^{(1)}$, a doubly degenerate TR breaking
state with $\Delta_{1}(\mathbf{k})=\Delta_{0}s_{z}(\lambda_{5}\pm i\lambda
_{7})$, (2) $E_{u}^{(2)}$, a TR invariant state with $\Delta_{2}%
(\mathbf{k})=\Delta_{0}i(s_{x}\lambda_{5}+s_{y}\lambda_{7})$, (Note that
$E_{u}^{(2^{\prime})}$ is equivalent to $E_{u}^{(2)}$), (3) $E_{u}^{(3)}$, a
doubly degenerate TR breaking state with $\Delta_{3}(\mathbf{k})=\Delta
_{0}[f_{3}(k_{x})\pm if_{3}(k_{y})]s_{z}\lambda_{0}$, (4) $A_{1u}^{(1)}$, a TR
invariant state with $\Delta_{4}(\mathbf{k})=$ $\Delta_{0}[f_{3}(k_{x}%
)s_{x}\lambda_{0}+f_{3}(k_{y})s_{y}\lambda_{0}]$, (Note that all four
components in \{$A_{1u}^{(1)}$: $f_{3}(k_{x})s_{x/y}\lambda_{0}\pm f_{3}%
(k_{y})s_{y/x}\lambda_{0}$\} are equivalent.). Through the bulk-boundary
correspondence, we demonstrate that all these four kinds of odd-parity pairing
channels support topological SC states in monolayer FeSe/STO. The BdG
Hamiltonian describing the SC states can be obtained by combining the
tight-binding Hamiltonian $H_{0}$ in Eq. (\ref{TB1}) and pairing term $H_{p}$
in Eq. (\ref{Hp}), \textit{i.e.},
\begin{equation}
H_{BdG}=H_{0}+H_{p}. \label{Hbdg}%
\end{equation}
Note that $H_{BdG}$ in Eq. (\ref{Hbdg}) includes both odd-orbital-parity and
even-orbital-parity parts. The edge spectra from the odd-orbital-parity parts
of $H_{BdG}$ with $\Delta_{1}(\mathbf{k})...\Delta_{4}(\mathbf{k})$ are
presented in Fig. 2. The even-orbital-parity parts of $H_{BdG}$ give the same
spectra if $k_{y}$ is translated to $k_{y}+\pi$ (see Fig.1(b) for comparison).
The edge spectra in Fig. 2 explicitly support the Andreev bound states which
are the identifications of topological superconductors. Besides, the bulk
properties of topological superconductors are usually characterized by some
topological numbers. Here, the pairing channels with $\Delta_{1}(\mathbf{k})$
and $\Delta_{3}(\mathbf{k})$ break the TR symmetry, and the Chern
number\cite{Thouless1982PRL} can be introduced to characterize such two
states, i.e., $\mathcal{C}=\frac{i}{2\pi}\sum_{E_{n}<0}%
%TCIMACRO{\dint \nolimits_{BZ}}%
%BeginExpansion
{\displaystyle\int\nolimits_{BZ}}
%EndExpansion
d\mathbf{k}\langle\nabla_{k}u_{n}(\mathbf{k})|\times|\nabla_{k}u_{n}%
(\mathbf{k})\rangle$. The calculations show that both odd-orbital-parity and
even-orbital-parity parts give the Chern numbers $\mathcal{C}^{o}=$
$\mathcal{C}^{e}=4$ in the one-Fe BZ for $\Delta_{1}(\mathbf{k})$ and
$\Delta_{3}(\mathbf{k})$ pairing channels. Thus, such two pairing channels are
characterized by the total Chern number $\mathcal{C=}\frac{1}{2}%
(\mathcal{C}^{o}+$ $\mathcal{C}^{e})=4$ in the two-Fe BZ. The Chern number
$\mathcal{C}=4$ is equal to the number of edge Andreev bound states shown in
Fig. 2 (a) and (d). For the TR invariant $\Delta_{2}(\mathbf{k})$ and
$\Delta_{4}(\mathbf{k})$ pairing channels, the total Chern numbers are zero.
However, the spin Chern numbers\cite{Sheng2006PRL,Wang2009PRB} can be
introduced to characterize the bulk topological properties of SC states in
$\Delta_{2}(\mathbf{k})$ or $\Delta_{4}(\mathbf{k})$ pairing channels. Namely,
$\mathcal{C}_{\uparrow}^{o/e}=1$, $\mathcal{C}_{\downarrow}^{o/e}=-1$ in the
two-Fe BZ. Correspondingly, two $Z_{2}$ topological numbers\cite{Kane2005PRL}
with opposite orbital parities defined by $v^{o/e}=\frac{1}{2}(\mathcal{C}%
_{\uparrow}^{o/e}-\mathcal{C}_{\downarrow}^{o/e})=1$ characterize the bulk
topological properties for SC states in $\Delta_{2}(\mathbf{k})$ or
$\Delta_{4}(\mathbf{k})$ pairing channels.\begin{figure}[ptb]
\begin{center}
\includegraphics[width=1\linewidth]{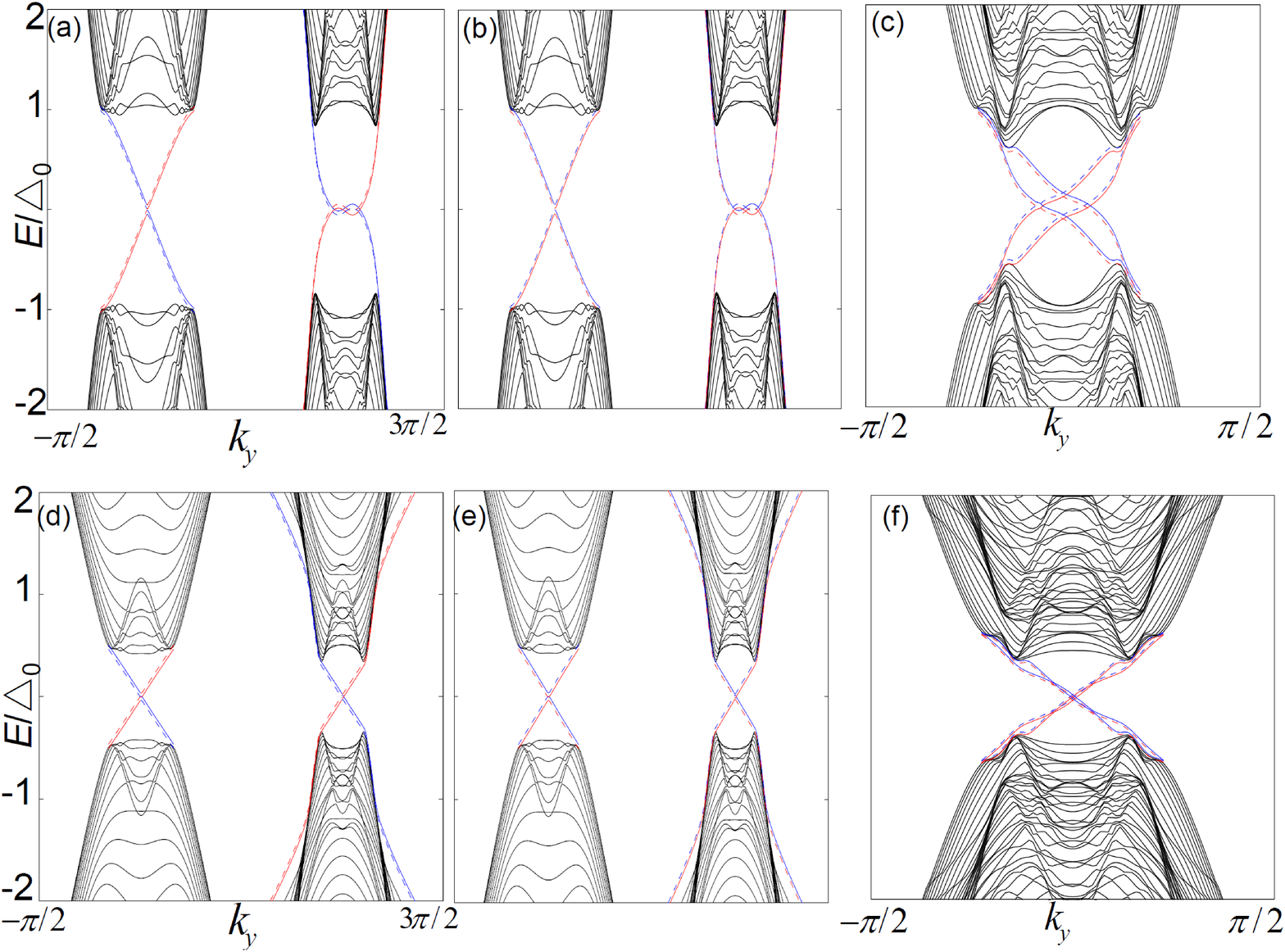}
\end{center}
\caption{The edge spectra of odd-orbital-parity BdG Hamiltonian with
$\Delta_{1}(k)$, $\Delta_{2}(k)$, $\Delta_{3}(k)$ and $\Delta_{4}(k)$ in (a),
(b), (d), and (e). In the presence of the orbital-parity-broken perturbation,
\textit{i.e.}, the staggered potential of Fe sublattices, the edge spectra of
BdG Hamiltonian with $\Delta_{2}(k)$ and $\Delta_{4}(k)$ are shown in (c) and
(f). Here, the system has periodic boundary condition along the $y$ direction
and open boundary condition along the $x$ direction with 51 one-Fe unit cell
lengths. The red/blue colors label the edge states localizing at the opposite
boundaries, and the dashed/solid lines label the edge states with up/down spin
directions. Note that the degenerate edge states on the same edge are
artificially split as a guide for the eye.}%
\end{figure}

Having confirmed that the topological SC states emerge in the nodeless
odd-parity pairing channels, we notice that the edge spectra shown in Fig. 2
(a) and (b) and the edge spectra shown in Fig. 2 (d) and (e) are very
different. Therefore, it is necessary to extract the minimum effective models
to clarify the essential physics hidden behind. First, we are aware of the
$\Delta_{3/4}(\mathbf{k})$ pairing channels being in the intra-orbital
spin-triplet $p$-wave pairing channels. Thus, the orbital degree of freedom is
inessential, and the minimum effective Hamiltonian can be reduced into the
single band space, which is the same Hamiltonian to describe the well-known
$p\pm ip$ topological
superconductors/superfluids\cite{Qi2011RMP,Read2000PRB,Qi2009PRL}, and the
nontrivial topology is referred to the $p\pm ip$ pairing terms. Therefore, we
omit our discussions for these \textquotedblleft trivial\textquotedblright%
\ topological SC states.

For $\Delta_{1}(\mathbf{k})$ and $\Delta_{2}(\mathbf{k})$, which are the
inter-orbital spin-triplet $s$-wave pairing channels, the three $t_{2g}$
orbitals are involved and entangled with each other not only in the bands
around the Fermi surface shown in Fig. 3 (a), but in the pairing terms shown
in Fig. 3(d). Note that we should have three bands when we consider three
$t_{2g}$ orbitals. It indicates that the third band mainly with the $d_{xz}$
and $d_{yz}$ weight has to strongly couple with two $e_{g}$ orbitals and be
gaped and pushed away from the Fermi level. In order to describe the two bands
in exact three orbital basis, we adopt the angular momentum representation
characterized by the azimuthal and magnetic quantum numbers $l$ and $m$. The
new electron creation operators are $d_{(lm=2,\pm1),\sigma}^{\dag}%
(\mathbf{k})=\mp\frac{1}{\sqrt{2}}[d_{xz,\sigma}^{\dag}(\mathbf{k})\pm
id_{yz,\sigma}^{\dag}(\mathbf{k})]$, then we have $\hat{\Delta}_{1}^{\dag
}(\mathbf{k})\sim\lbrack d_{(2,1),\uparrow}^{\dag}(\mathbf{k})d_{xy,\downarrow
}^{\dag}(-\mathbf{k})+d_{(2,1),\downarrow}^{\dag}(\mathbf{k})d_{xy,\uparrow
}^{\dag}(-\mathbf{k})]$ and $\hat{\Delta}_{2}^{\dag}(\mathbf{k})\sim\lbrack
d_{(2,-1),\uparrow}^{\dag}(\mathbf{k})d_{xy,\uparrow}^{\dag}(-\mathbf{k}%
)+d_{(2,1),\downarrow}^{\dag}(\mathbf{k})d_{xy,\downarrow}^{\dag}%
(-\mathbf{k})]$. Now, we can only exploit the operators involving in
$\hat{\Delta}_{1/2}(\mathbf{k})$ to construct the basis to write the minimum
effective Hamiltonian, and this approximation is equivalent to treating with
$d_{xz}$ and $d_{yz}$ orbitals with equal weights. In the effective basis,
$\Psi_{1/2}(\mathbf{k})=[\{\psi_{1/2\uparrow}(\mathbf{k})\}$, $\{\psi
_{1/2\downarrow}(\mathbf{k})\}]^{T}$ with $\{\psi_{1/2,\sigma}(\mathbf{k}%
)\}=\{d_{(2,1/-(-1)^{\sigma}),\sigma}(\mathbf{k})$, $d_{xy,\sigma}%
(\mathbf{k})$, $d_{xy,\bar{\sigma}/\sigma}^{\dag}(-\mathbf{k})$,
$-d_{(2,1/-(-1)^{\sigma})\bar{\sigma}}^{\dag}(-\mathbf{k})\}$,
\begin{equation}
H^{(1/2)}(\mathbf{k})=\mathcal{H}_{1}^{(1/2)}(\mathbf{k})\oplus\mathcal{H}%
_{2}^{(1/2)}(\mathbf{k}). \label{Heff1}%
\end{equation}
Here, $\mathbf{k}$ is measured from the $M$ point. $\bar{\sigma}=-\sigma$, and
$(-1)^{\sigma}=1$/$-1$ for spin $\downarrow$/$\uparrow$, the orbital parity
index is omitted for simplicity. $\mathcal{H}_{1}^{(1/2)}(\mathbf{k})=$
$\tau_{z}[d_{0}^{(1/2)}(\mathbf{k})+\sum_{i=x}^{z}d_{i}^{(1/2)}(\mathbf{k}%
)\sigma_{i}]+\tau_{x}\Delta_{0}$, $\mathcal{H}_{2}^{(1)}(\mathbf{k}%
)=\mathcal{H}_{1}^{(1)}(\mathbf{k})$ and $\mathcal{H}_{2}^{(2)}(\mathbf{k}%
)=\mathcal{H}_{1}^{(2)\ast}(-\mathbf{k})$. The three Pauli matrices
$\sigma_{1/2/3}$ are introduced to span the effective two-band space.
$d_{0}^{(1/2)}(\mathbf{k})=\frac{\varepsilon_{1}(\mathbf{k})+\varepsilon
_{2}(\mathbf{k})}{2}-\mu$, $d_{x}^{(1/2)}(\mathbf{k})=\mp Ak_{y}$,
$d_{y}^{(1/2)}(\mathbf{k})=-Ak_{x}$ and $d_{z}^{(1/2)}(\mathbf{k}%
)=\frac{\varepsilon_{1}(\mathbf{k})-\varepsilon_{2}(\mathbf{k})}{2}$.
$H^{(1)}(\mathbf{k})$ breaks TR symmetry, because only $m=1$ is involved.
$H^{(2)}(\mathbf{k})$ is TR invariant, and characterized by the $T^{-1}%
H^{(2)}(\mathbf{k})T=$ $H^{(2)^{\ast}}(-\mathbf{k})$, where the TR symmetry
operator is $T=is_{y}\tau_{0}\sigma_{0}\mathcal{K}$ with $\mathcal{K}$ the
complex conjugated operator. The dispersions $\varepsilon_{1/2}(\mathbf{k})$
with definite orbital parity can be read out from Fig. 1 (e) and Fig. 3 (b).
Around $M_{y}$ point, we have $\varepsilon_{1/2}^{e}(\mathbf{k})=e_{1/2}%
-\mu+\alpha_{1/2}k_{x}^{2}+\beta_{1/2}k_{y}^{2}$ and $\varepsilon_{1/2}%
^{o}(\mathbf{k})=e_{1/2}-\mu+\beta_{1/2}k_{x}^{2}+\alpha_{1/2}k_{y}^{2}$. The
signs of $\alpha$/$\beta$ are crucial to determine the properties of the
topological SC states. In Fig. 3 (b) and (c), we schematically illustrate the
evolution of the $\varepsilon_{1/2}^{o}(\mathbf{k})$ under the couplings
induced by glide plane around $M_{y}$ point, and we can find $e_{1}<e_{2}$,
$\alpha_{1}<0$, $\beta_{1}>0$, $\alpha_{2}>0$, $\beta_{2}<0$. The effective
mass measuring the energy gap $E_{M_{3}}-E_{M_{1}}$ shown in Fig. 1 (e) or
Fig. 3 (c) is $m=\frac{e_{2}-e_{1}}{2}>0$. The finite electron-doped condition
$\mu^{2}+\Delta_{0}^{2}>m^{2}$\cite{Fu2012PRL-1} always supports topological
SC states for $\mathcal{H}_{1}^{(1/2)}(\mathbf{k})$, where the chemical
potential $\mu$ is measured from the middle of gap. The remarkable feature of
the edge spectra in Fig. 2 (a) and (b) is that the edge Andreev bound states
have a twist (three times of crossings) around $k_{y}=\pi$ and only one
crossing around $k_{y}=0$. This difference can be understood with the
\textquotedblleft orbital mirror helicity\textquotedblright\ from the mirror
operator in $c_{2}(x/y)$ acting on three $t_{2g}$ orbitals in analogy to the
\textquotedblleft spin mirror helicity\textquotedblright\ proposed in
Ref\cite{Fu2012PRL-1}. The conservation of mirror helicity force the
non-/twisted feature of the edge Andreev edge states under the
non-/band-inversion conditions between $\varepsilon_{1}^{e/o}(\mathbf{k})$ and
$\varepsilon_{2}^{e/o}(\mathbf{k})$ along $x$ direction, $\mathrm{sgn}%
[(e_{2}-e_{1})(\alpha_{2}-\alpha_{1})]>0$/$\mathrm{sgn}[(e_{2}-e_{1}%
)(\beta_{2}-\beta_{1})]<0$ (Note that $\varepsilon_{1/2}^{e}(\mathbf{M}%
_{y}+\mathbf{k})=\varepsilon_{1/2}^{o}(\mathbf{M}_{x}+\mathbf{k})$). We are
aware of the importance of the non-symmorphic lattice symmetry which not only
induces the orbital-momentum locking $\mathbf{k}\times\mathbf{\sigma\cdot
\hat{z}}$ through glide plane, but protects the exotic behaviors of the edge
Andreev bound states. We can verify this point through introducing the
staggered onsite potential, which mixes the orbital parities, breaks the
non-symmorphic lattice symmetry and destroys the twist feature of edge
spectra. The results are shown in Fig. 2 (c) and (f). However, the bulk
topological properties are robust against such perturbations.
\begin{figure}[ptb]
\begin{center}
\includegraphics[width=1\linewidth]{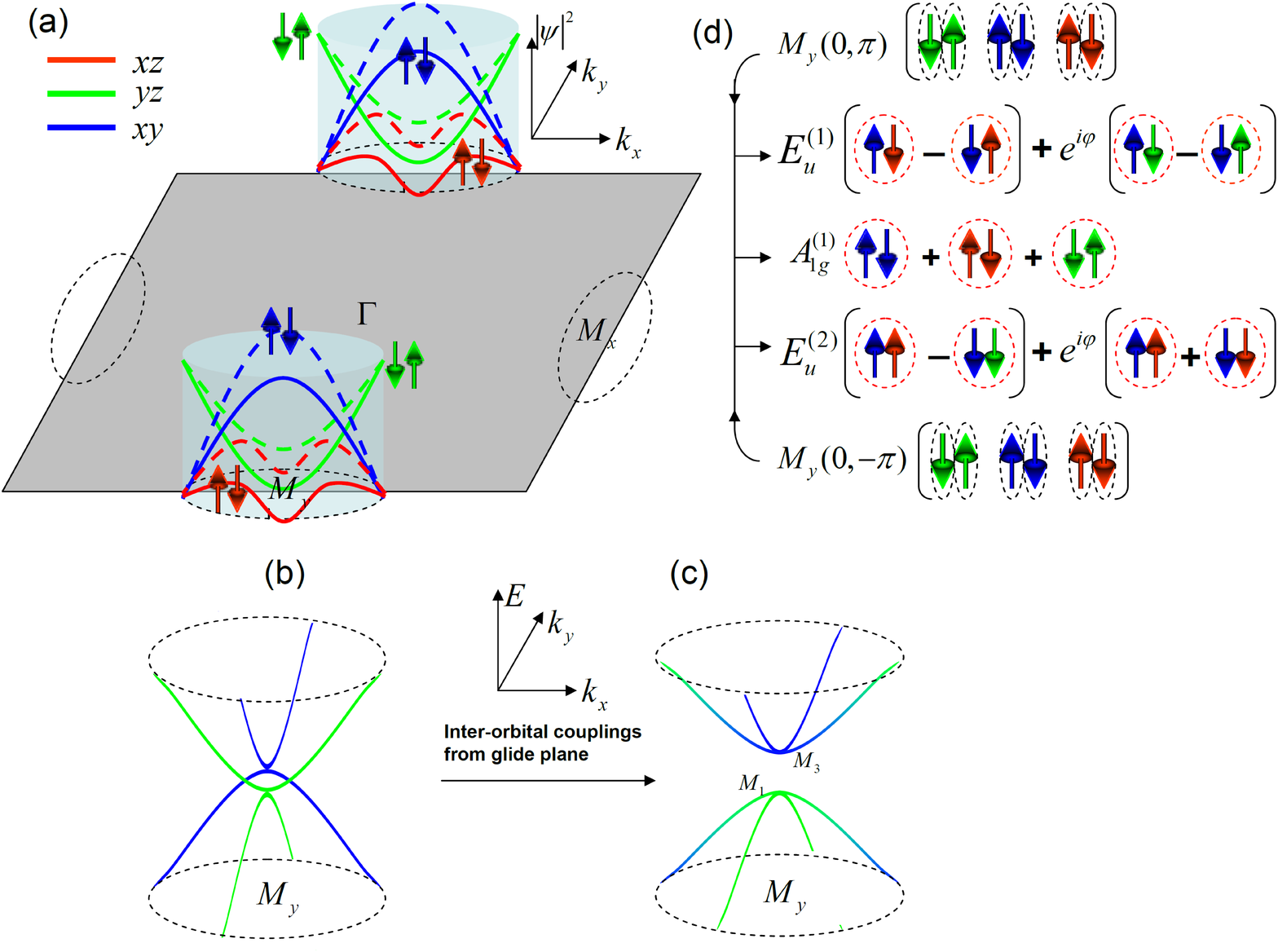}
\end{center}
\caption{(a) The weight of three $t_{2g}$ along the Fermi surface around
$M_{y}$ with odd-orbital parity. (b) and (c) The effective band dispersions
without/with inter-orbital coupling from glide plane. (d) Three competitive
pairing channels with $\varphi=\frac{\pi}{2}$in weak-coupling limit.}%
\end{figure}

\section{The effective pairing interactions}

Although the high temperature interfacial superconductivity in monolayer
FeSe/STO seems to have been established beyond doubt, the mechanism for
superconductivity is still an open question\cite{Bozovic2014NP}, and the
unique features of monolayer FeSe/STO further pose a higher barrier to block
our understanding of the superconductivity from some standard theories. For
example, the monolayer FeSe/STO is strictly two-dimensional and has no hole
pockets at the BZ center while its three-dimensional counterpart bulk FeSe
resembles iron-pnictide with hole pockets. The Fermi surface of monolayer
FeSe/STO is similar to that of A$_{x}$Fe$_{2-y}$Se$_{2}$ (A=K, Cs, Rb), except
that the small electron pocket around $(0,0,\pi)$ in A$_{x}$Fe$_{2-y}$Se$_{2}$
is absent here. In weak coupling limit, the spin-fluctuation-exchange theory
predicts that the \{$B_{1g}$: $f_{2}(k)s_{0}\lambda_{0}$\} pairing channel is
dominant in A$_{x}$Fe$_{2-y}$Se$_{2}$ and the gap structure has nodes along
the $k_{z}$ direction\cite{Maier2011PRB,Kreisel2013PRB}. However, the ARPES
measurements reported isotropic full gaps without nodes on all pockets in
A$_{x}$Fe$_{2-y}$Se$_{2}$\cite{ZhangNM2011,XuPRB2012}. In the strong coupling
limit, the phenomenological $t$-$J$ model predicts that the \{$A_{1g}$:
$f_{4}(k)s_{0}\lambda_{0}$\} pairing channel is dominant in A$_{x}$Fe$_{2-y}%
$Se$_{2}$ and the gaps have same sign for all the pockets\cite{Fang2011PRX}.
However, the inelastic neutron scattering measurements on A$_{x}$Fe$_{2-y}%
$Se$_{2}$ reported a resonance with wave vector $\mathbf{Q}_{c}=(\pi,\pi/2)$
in the superconducting state\cite{Park2011PRL}, which indicated that there
existed a sign-change between the fermi surfaces connected by $\mathbf{Q}_{c}%
$. These contradictions strongly question the standard theories. On the other
hand, the studies of some confirmed systems with interfacial superconductivity
including bilayer lanthanum cuprate\cite{Bozovic2002PRL} and LaAlO$_{3}%
$/SrTiO$_{3}$ heterostructure\cite{Reyren2007S} could provide us some useful
insights to understand the superconductivity in mono-layer FeSe/STO. The
studies of the aforementioned systems indicate that surface phonon plays a key
role to drive the superconductivity\cite{Ginzburg1964PL}. Recent ARPES
experiment observed the band replication, which was attributed to strong
coupling between the cross phonon and electrons\cite{Lee2014N}, and the
cooperation between the cross phonon mode and spin fluctuation is argued to be
the origin to enhance $T_{c}$ in monolayer FeSe/STO. Therefore, it is still
possible that the superconductivity in monolayer FeSe/STO is driven by the
electron-phonon coupling, and the surface phonon-mediated SC mechanism in
monolayer FeSe/STO has been proposed in Ref.\cite{Rademaker2015Arxiv}. Here,
without loss of generality, we consider several possibilities of the effective
interactions that can drive superconductivity in different pairing channels
and focus on the parameter regime missed previously.

We first assume the multi-orbital Hubbard interactions as pairing driver,%

\begin{align}
H_{int}^{(1)} &  =U\sum_{i,l}n_{il\uparrow}n_{il\downarrow}+V\sum
_{i,l>l^{\prime}}n_{il}n_{il^{\prime}}\nonumber\\
&  -J_{H}\sum_{i,l>l^{\prime}}(2\mathbf{S}_{il}\cdot\mathbf{S}_{il^{\prime}%
}+\frac{1}{2}n_{il}n_{il^{\prime}})\nonumber\\
&  +J^{\prime}\sum_{i,l\neq l^{\prime}}d_{i,l\uparrow}^{\dag}d_{i,l\downarrow
}^{\dag}d_{i,l^{\prime}\downarrow}d_{i,l^{\prime}\uparrow}.\label{H_int1}%
\end{align}
Here, $U$, $V$, $J_{H}$, $J^{\prime}$ are the intra-orbital, inter-orbital,
Hund's coupling and pairing hopping term. $l,l^{\prime}\in(xz,yz,xy)$, and
$\mathbf{S}_{il}=\frac{1}{2}d_{il\sigma}^{\dag}\mathbf{s}_{\sigma
\sigma^{\prime}}d_{il\sigma}$. The spin rotation symmetry requires
$U=V+2J_{H}$, and $J_{H}=J^{\prime}$ at the atomic level. Since the
predictions from the weak-coupling theory\cite{Maier2011PRB,Kreisel2013PRB}
about $H_{0}+$ $H_{int}^{(1)}$ were not consistent with the experimental
reports\cite{ZhangNM2011,XuPRB2012}, the strongly correlative picture with
quite large $J_{H}$ is possible and the strongly correlative effects in iron
chalcogenides have been reported by recent ARPES experiments\cite{Yi2015Arxiv}%
. Define the pairing operators,%

\[
\hat{\Delta}_{s,ll^{\prime}}=\sum_{k}\hat{\Delta}_{s,ll^{\prime}}%
(k),\hat{\Delta}_{t,ll^{\prime}}^{\alpha}=\sum_{k}\hat{\Delta}_{t,ll^{\prime}%
}^{\alpha}(k),
\]

\begin{align}
\hat{\Delta}_{s,ll^{\prime}}(k)  &  =\sum_{\sigma\sigma^{\prime}}\frac
{[is_{y}]_{\sigma\sigma^{\prime}}}{4}[d_{l\sigma}(\mathbf{k})d_{l^{\prime
}\sigma^{\prime}}(-\mathbf{k})+d_{l^{\prime}\sigma}(\mathbf{k})d_{l\sigma
^{\prime}}(-\mathbf{k})],\nonumber\\
\hat{\Delta}_{t,ll^{\prime}}^{\alpha}(k)  &  =\sum_{\sigma\sigma^{\prime}%
}\frac{[is_{y}s_{\alpha}]_{\sigma\sigma^{\prime}}}{4}[d_{l\sigma}%
(\mathbf{k})d_{l^{\prime}\sigma^{\prime}}(-\mathbf{k})-d_{l^{\prime}\sigma
}(\mathbf{k})d_{l\sigma^{\prime}}(-\mathbf{k})]. \label{pair_mf}%
\end{align}
The interaction Hamiltonian has the form,
\begin{align}
H_{int}^{(1)}  &  =U\sum_{l}\hat{\Delta}_{s,ll}^{\dag}\hat{\Delta}%
_{s,ll}+J_{H}\sum_{l\neq l^{\prime}}\hat{\Delta}_{s,ll}^{\dag}\hat{\Delta
}_{s,l^{\prime}l^{\prime}}\nonumber\\
&  +(V-J_{H})\sum_{ll^{\prime}\alpha}\hat{\Delta}_{t,ll^{\prime}}^{\alpha\dag
}\hat{\Delta}_{t,ll^{\prime}}^{\alpha}+(V+J_{H})\sum_{l\neq l^{\prime}}%
\hat{\Delta}_{s,ll^{\prime}}^{\dag}\hat{\Delta}_{s,ll^{\prime}}.
\label{H_int1c}%
\end{align}
When the Hund's coupling is strong enough, i.e., $J_{H}>U/3$, the third term
of Eq.(\ref{H_int1c}) can give rise to the instability in a spin-triplet
channel\cite{Lee2008PRB,Puetter2012EPL}, which involving the \{$A_{1g}$:
$is_{z}\lambda_{2}$\}, $E_{u}^{(1)}$ and $E_{u}^{(2)}$ IRs in Table
\ref{pairk-k}. The detailed discussions about these pairing channels are
merged into the third kind of effective interaction in the following.

Another standard theory for the superconductivity is the phenomenological
Heisenberg model in strong coupling limit, we consider the effectively
frustrated Heisenberg interaction\cite{Anderson1987S} as the pairing force,
\begin{equation}
H_{int}^{(2)}=J_{1}\sum_{l,<i,j>}\mathbf{S}_{il}\cdot\mathbf{S}_{jl}+J_{2}%
\sum_{l,\ll i,j\gg}\mathbf{S}_{il}\cdot\mathbf{S}_{jl}. \label{Hint2}%
\end{equation}
Here, $J_{1/2}$ are the nearest and next nearest neighbor magnetic exchange
couplings. A well-know result of $H_{int}^{(2)}$ is that the magnetic ground
state is checkerboard-antiferromagnetic when $2J_{2}<\left\vert J_{1}%
\right\vert $, and collinear-antiferromagnetic when $2J_{2}>\left\vert
J_{1}\right\vert $. However, no Fermi surface reconstruction induced by spin
density wave was observed in monolayer FeSe/STO but in mutli-layer FeSe/STO in
ARPES experiments\cite{Tan2013NM}. The recent first-principles calculations
proposed that the magnetic order was strongly frustrated in monolayer FeSe/STO
with \ $2J_{2}\approx\left\vert J_{1}\right\vert $\cite{Liu2015PRB}. Another
issue is the sign of $J_{1}$. If both $J_{1}$ and $J_{2}$ are
antiferromagnetic, the $\Delta_{3/4}(\mathbf{k})$ pairing channels are ruled
out, and the SC states fall into \{$A_{1g}$: $f_{4}(k)s_{0}\lambda_{0}$\}
induced by $J_{2}$ or \{$B_{1g}$: $f_{2}(k)s_{0}\lambda_{0}$\} induced by
$J_{1}$. If $J_{1}$ is ferromagnetic and $J_{2}$ are antiferromagnetic, the
$\Delta_{3/4}(\mathbf{k})$ pairing channels are possible from symmetry point,
but these two odd-parity pairing channels have to compete with the \{$A_{1g}$:
$f_{4}(k)s_{0}\lambda_{0}$\} induced by $J_{2}$. The winner is determined by
the topology of the Fermi surface\cite{Hu2012SR}. For the low electron-doped
at 0.1e/Fe, the Fermi pockets locating at $M$ points are quite small.
Therefore, the form factor $f_{4}(k)$ has large magnitude, and the SC states
favor the \{$A_{1g}$: $f_{4}(k)s_{0}\lambda_{0}$\}. If the electron-doped
level can be tuned in monolayer FeSe/STO without suppressing the
superconductivity. We can expect that the SC states in over electron-doped
samples would favor $\Delta_{3/4}(\mathbf{k})$ pairing channels for
ferromagnetic $J_{1}$, because the Fermi surface locates at the $X$ points,
where the form factors $f_{3}(k_{x/y})$ have large magnitudes. We note that
such kind of pairing was discussed in underdoped cuprates\cite{Lu2014NP}.

From the aforementioned arguments about the possibly significant role of
surface phonon, we consider the third kind of phenomenological interaction
from phonon-mediated mechanism\cite{Rademaker2015Arxiv} to induce the
interfacial SC instability in monolayer FeSe/STO,
\begin{equation}
H_{int}^{(3)}=\underset{l,l^{\prime},\sigma,\sigma^{\prime},\mathbf{k}%
,\mathbf{k}^{\prime}}{\sum}\frac{1}{2}V_{l,l^{\prime}}^{\sigma,\sigma^{\prime
}}(\mathbf{k},\mathbf{k}^{\prime})d_{k,l\sigma}^{\dag}d_{-k,l^{\prime}%
\sigma^{\prime}}^{\dag}d_{-k^{\prime}l^{\prime}\sigma^{\prime}}d_{k^{\prime
},l\sigma}. \label{H_int3}%
\end{equation}
Here, we assume $V_{l,l^{\prime}}^{\sigma,\sigma^{\prime}}(\mathbf{k}%
,\mathbf{k}^{\prime})=-V_{0}$ for $l=l^{\prime}$, $\sigma^{\prime}=\bar
{\sigma}$ and $V_{l,l^{\prime}}^{\sigma,\sigma^{\prime}}(\mathbf{k}%
,\mathbf{k}^{\prime})=-V_{1}$ for $l>l^{\prime}$. Note that the third term in
Eq.(\ref{H_int1c}) with $J_{H}>U/3$ can also be described by $H_{int}^{(3)}$.
With the pairing operators shown in Eq. (\ref{pair_mf}), $H_{int}^{(3)}$ takes
the form,%

\begin{align}
H_{int}^{(3)}  &  =-V_{0}\sum_{l}\hat{\Delta}_{s,ll}^{\dag}\hat{\Delta}%
_{s,ll}-V_{1}\sum_{l>l^{\prime}}\hat{\Delta}_{s,ll^{\prime}}^{\dag}\hat
{\Delta}_{s,ll^{\prime}}\nonumber\\
&  -V_{1}\sum_{l>l^{\prime}\alpha}\hat{\Delta}_{t,ll^{\prime}}^{\alpha\dag
}\hat{\Delta}_{t,ll^{\prime}}^{\alpha}. \label{H_int33}%
\end{align}
Under the mean-field approximation, $\Delta_{s,ll^{\prime}}=\langle\hat
{\Delta}_{s,ll^{\prime}}^{\dag}\rangle$, $\Delta_{t,ll^{\prime}}^{\alpha
}=\langle\hat{\Delta}_{t,ll^{\prime}}^{\alpha}\rangle$, the $H_{int}^{(3)}$
can be decoupled as follows,
\begin{align}
H_{int}^{(3)}  &  =-V_{0}\sum_{l}\Delta_{s,ll}\hat{\Delta}_{s,ll}^{\dag}%
-V_{1}\sum_{l>l^{\prime}}\Delta_{s,ll^{\prime}}\hat{\Delta}_{s,ll^{\prime}%
}^{\dag}\nonumber\\
&  -V_{1}\sum_{l>l^{\prime}\alpha}\Delta_{t,ll^{\prime}}^{\alpha}\hat{\Delta
}_{t,ll^{\prime}}^{\alpha\dag}+H.c.+h_{con}. \label{Hmfnew}%
\end{align}
Here $h_{con}=\sum_{l}V_{0}|\Delta_{s,ll}|^{2}+V_{1}\sum_{l>l^{\prime}}%
|\Delta_{s,ll^{\prime}}|^{2}+V_{1}\sum_{l>l^{\prime},\alpha}|\Delta
_{s,ll^{\prime}}^{\alpha}|^{2}$. Now, we consider the odd-orbital-parity parts
of the normal-state Hamiltonian. The mean-field Hamiltonian takes the
following form,%

\begin{equation}
H_{MF}=\sum_{k}\frac{1}{2}\Psi^{\dag}(k)H_{MF}(k)\Psi(k)+H_{con},
\label{Hmf12}%
\end{equation}
where $\Psi(k)$ has the same form shown in Eq.(\ref{Hp}) except $\{d_{\sigma
}(\mathbf{k})\}=\{d_{xz,\sigma}(\mathbf{k}),d_{yz,\sigma}(\mathbf{k}%
),d_{xy,\sigma}(\mathbf{k}),d_{x^{2}-y^{2},\sigma}(\mathbf{k}),d_{z^{2}%
,\sigma}(\mathbf{k})\}$ now. Then $H_{MF}(k)=H_{0}(k)\tau_{z}+\Delta
(k)\tau_{x}$, $H_{0}(k)$ $=A_{o}(k)\oplus A_{o}(k)$, and $H_{con}%
=\underset{k,m=1}{\overset{5}{\sum}}A_{o,mm}(k)+h_{con}$. Assume the
$H_{MF}(k)$ can be diagonalized with matrix $\tilde{U}_{k}$, i.e., $\tilde
{U}_{k}^{\dag}H_{MF}(k)\tilde{U}_{k}=E_{k,1}\oplus E_{k,2}...E_{k,20}$. Then
the mean-field self-consistent equations take the forms,%

\begin{align}
\Delta_{s,ll^{\prime}}  &  =\sum_{k,n=1}^{20}\frac{[\tilde{U}_{k,n,l}^{\ast
}\tilde{U}_{k,n,l^{\prime}+10}+\tilde{U}_{k,n,l+5}^{\ast}\tilde{U}%
_{k,n,l^{\prime}+15}]f(E_{k,n})}{2},\nonumber\\
\Delta_{t,ll^{\prime}}^{x}  &  =\sum_{k,n=1}^{20}\frac{-[\tilde{U}%
_{k,n,l}^{\ast}\tilde{U}_{k,n,l^{\prime}+15}+\tilde{U}_{k,n,l+5}^{\ast}%
\tilde{U}_{k,n,l^{\prime}+10}]f(E_{k,n})}{2},\nonumber\\
\Delta_{t,ll^{\prime}}^{y}  &  =\sum_{k,n=1}^{20}\frac{-i[\tilde{U}%
_{k,n,l}^{\ast}\tilde{U}_{k,n,l^{\prime}+15}-\tilde{U}_{k,n,l+5}^{\ast}%
\tilde{U}_{k,n,l^{\prime}+10}]f(E_{k,n})}{2},\nonumber\\
\Delta_{t,ll^{\prime}}^{z}  &  =\sum_{k,n=1}^{20}\frac{-[\tilde{U}%
_{k,n,l}^{\ast}\tilde{U}_{k,n,l^{\prime}+10}-\tilde{U}_{k,n,l+5}^{\ast}%
\tilde{U}_{k,n,l^{\prime}+10}]f(E_{k,n})}{2},\nonumber\\
N_{e}  &  =\sum_{k,n=1}^{20}\sum_{m=1}^{10}|\tilde{U}_{k,n,m}^{\ast}%
|^{2}f(E_{k,n}). \label{selfeq}%
\end{align}
Here, $f(x)=\frac{1}{e^{\frac{x}{k_{B}T}}+1}$ is the Fermi distribution
function and $N_{e}$ is the electron number. In comparison with Table
\ref{pairk-k} and Eq.(\ref{Hmfnew}), the relevant IR channels in Table
\ref{pairk-k} can be represented with (\ref{selfeq}). For example,
\{$A_{1g}^{(1)}$: $s_{0}\lambda_{0}$\}$=s_{0}(\Delta_{s,xz,xz}\oplus
\Delta_{s,yz,yz}\oplus\Delta_{s,xy,xy})$, \{$E_{u}^{(2)}$: $i(s_{x}\lambda
_{5},s_{y}\lambda_{7})$\}$=i(\Delta_{t,xz,xy}^{x}s_{x}\lambda_{5}%
,\Delta_{t,xz,xy}^{y}s_{y}\lambda_{7})$. Likewise, other IR channels can be
read out following same way.

It is possible for $\Delta(k)$ to take the form of linear combinations of
several different IR channels, but some symmetries have to be broken to pay
the price for such coexistence. For example the inverse symmetry is broken for
the SC states proposed in Refs.\cite{Hu2013PRX,Hao2014PRB}. Likewise, the TR
symmetry or lattice symmetry could also be broken when two different
one-dimensional IRs or two components in a two-dimensional IR coexist. In
order to gain some insight before we perform the numerical calculations, we
note that all the experiments reported the isotropic Fermi surface and gap
structures without any resolvable distortions, and the monolayer FeSe/STO was
conformed to be the cleanest composition with the simplest
structure\cite{Liu2012NC,He2013NM,Tan2013NM}. These features rule out the
possibilities of some complex orders, such as nematic order found in bulk
FeSe. From Table \ref{pairk-k}, we can first eliminate the possibilities of
the \{$B_{2g}$: $s_{0}\lambda_{1}$\}, \{$A_{1g}$: $is_{z}\lambda_{2}$\},
\{$E_{g}$: $i(s_{x},s_{y})\lambda_{2}$\} and \{$A_{1g}$: $s_{0}\lambda_{8}$\}
pairing channels, because the leading inter-$d_{xz}$-$d_{yz}$ hopping term is
proportional to $\sin k_{x}\sin k_{y}$, which is nearly zero around the Fermi
surface, and the \{$A_{1g}$: $s_{0}\lambda_{8}$\} channel has nodes. Second,
it is straightforward to check that two components in \{$E_{u}$:
$s_{0}(\lambda_{4},\lambda_{6})$\} or \{$E_{u}^{(1)}$: $s_{z}(\lambda
_{5},\lambda_{7})$\} give two degenerate strip SC states with nodes. Thus, the
TR-broken linear combination of two components is optimal to achieve the
isotropic nodeless gap structure and lower the energy. Note that the
coexistence of these two two-dimensional IRs could rise the energy, because
they follow different transformations under the lattice symmetric operations
and suppress the gap amplitude. Finally, no additionally global symmetries can
be broken for \{$A_{1g}^{(1)}$: $s_{0}\lambda_{0}$\} and \{$E_{u}^{(2)}$:
$i(s_{x}\lambda_{5},s_{y}\lambda_{7})$\} to coexist with each other and with
\{$E_{u}$: $s_{0}(\lambda_{4},\lambda_{6})$\} or \{$E_{u}^{(1)}$:
$s_{z}(\lambda_{5},\lambda_{7})$\} to avoid breaking the isotropic SC gap
structure and achieving lower energy. Therefore, we find that these four IRs,
i.e., \{$E_{u}$: $s_{0}(\lambda_{4},\lambda_{6})$\}, \{$E_{u}^{(1)}$:
$s_{z}(\lambda_{5},\lambda_{7})$\}, \{$A_{1g}^{(1)}$: $s_{0}\lambda_{0}$\} and
\{$E_{u}^{(2)}$: $i(s_{x}\lambda_{5},s_{y}\lambda_{7})$\} are independent, and
TR symmetry should be spontaneously broken in the first two IRs. It is
straightforward to verify these arguments through the following numerical
calculations.\begin{figure}[ptb]
\begin{center}
\includegraphics[width=1\linewidth]{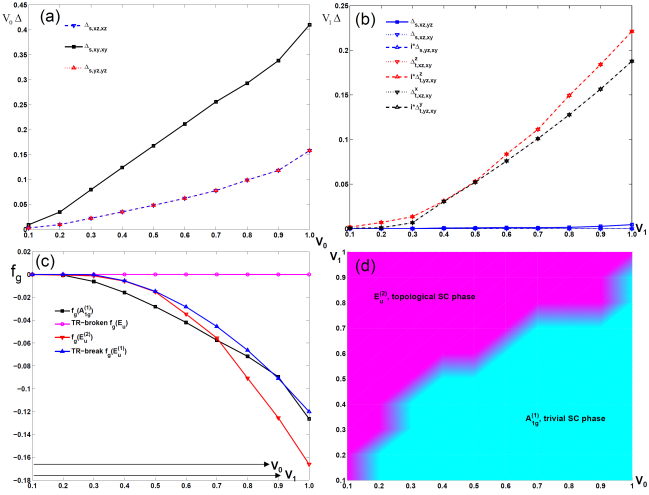}
\end{center}
\caption{(a) The evolution of three components of SC order parameters in
$A_{1g}^{(1)}$ channel about $V_{0}$. (b) The evolution of components of SC
order parameters in $E_{u}$, $E_{u}^{(1)}$ and $E_{u}^{(2)}$ channels about
$V_{1}$. (c) The evolution of the condensed energy in different SC states with
relevant IRs about $V_{0}$ and $V_{1}$. (d) The phase diagram is plotted in
($V_{0},$ $V_{1}$) plane with respect to the lowest energy. We set 51$\times
$51mesh of $\mathbf{k}$, and the electron number to satisfy electron-doped
0.1e/Fe. The energy scale is measured with eV.}%
\end{figure}

Now, we perform the numerical calculations to evaluate which pairing channel
governs the ground state of the system for different $V_{0}$ and $V_{1}$. The
ground state energy of Eq.(\ref{Hmf12}) is $G_{s}(T)=-k_{B}T\ln$Tr$e^{-\beta
H_{MF}}$, and $G_{s}(T\sim0)=(H_{con}-\frac{1}{2}\underset{k,n=1}{\overset
{10}{\sum}}|E_{k,n}|)$ at zero temperature. For simplicity, we can evaluate
the ground state through the minimum of the condensed energy density defined
as $f_{g}=h_{con}-\frac{1}{8\pi^{2}}\underset{n=1}{\overset{10}{\sum}}\int
d^{2}\mathbf{k}|E_{k,n}|-\frac{1}{4\pi^{2}}\underset{n=1}{\overset{5}{\sum}%
}\int d^{2}\mathbf{k}|E_{k,n}^{o}|$ for given electron number, where
$E_{k,n}^{o}$ are the energy spectra of normal state. Solve the
self-consistent equations (\ref{Hmf12}) and (\ref{selfeq}) for parameters
$(V_{0},V_{1})$ with respect to the minimum of $f_{g}$, we show the evolution
of SC order parameters and condensed energy about $(V_{0},V_{1})$ in Fig. 4,
and we find topologically trivial \{$A_{1g}^{(1)}$: $s_{0}\lambda_{0}$\}
channel and topologically nontrivial \{$E_{u}^{(2)}$: $i(s_{x}\lambda
_{5},s_{y}\lambda_{7})$\} are dominant in relevant regime of $(V_{0},V_{1})$
parameter plane.

\section{Discussion and summary}

If the superconductivity in monolayer FeSe/STO is driven by the effective
interaction $H_{int}^{(3)}$ in Eq. (\ref{H_int33}), the observed isotropic and
nodeless s-wave gap structures select both topologically trivial $A_{1g}%
^{(1)}$ ($s_{0}\lambda_{0}$) and nontrivial $E_{u}^{(2)}$($\Delta
_{2}(\mathbf{k})$) as possible candidates. The essential difference lies in
that the former one has even-parity and spin-singlet pairing while the latter
one has odd-parity and spin-triplet pairing. Therefore, it is unambiguous to
adopt the experiments which can directly distinguish the spin states and
parities to pin down the possible candidate. Particularly, temperature
dependence of the nuclear magnetic relaxation(NMR) rate can be utilized to
distinguish the two different pairings. The well-known result is that the NMR
rate has a Hebel-Slichter peak at the SC transition temperature for the
even-parity and spin-singlet s-wave SC state\cite{Hebel1959PR}. However, the
Hebel-Slichter peak could disappear with the anti-peak behavior due to the
unique spin, orbital, and momentum locking effect in topological SC states
with odd-parity as shown in Ref.\cite{Nagai2015Arxiv}. The parity of the
Cooper pair is characterized by the inverse operator $\{i|\frac{1}{2}\frac
{1}{2}\}$. It indicates the odd-parity pairing has a sign change or phase
shift of $\pi$ between the top Se and and bottom Se layers along the $c$-axis
compared with the even-parity pairing. Thus, the standard magnetic-flux
modulation of dc SC quantum interference devices (SQUIDS)
measurements\cite{Wollman1993PRL,Hu2012PRX,Fu2010PRL} provide another scheme
to distinguish the odd- and even-parity pairings. On the other hand, some
transport measurements can also be applied to detect the topological
superconductors, such as the thermal Hall
conductivity\cite{Shiozaki2013PRL,Yosuke2014arXiv}. The challenge for such
measurements is that the FeSe is very air sensitive, and the experimental
measurements should be performed under the ultra-high vacuum condition.

In the aforementioned discussions about the SC pairings, we assume that the
glide-plane symmetry is not broken. Actually, there exist some possible
effects to break the glide-plane symmetry. For example, the atomic
spin-orbital coupling could be have non-neglectable effect in iron
chalcogenides. It is explicit that the inter-orbital spin-orbital coupling can
mix the bands with inverse orbital parities, and induce the inter-orbital SC
pairing in $(\mathbf{k},-\mathbf{k}+\mathbf{Q})$ channels. However, the weight
of inter-$d_{xz}$-$d_{xy}$ spin-orbital coupling is proportional to
$\lambda_{so}\sim0.05eV$\cite{Kreisel2013PRB}, while the inter-$d_{xz}%
$-$d_{xy}$ orbital hopping term with definite orbital parity is proportional
to $|2it_{x}^{14}\sin k_{F}|$ $\sim0.3eV$ at the Fermi surface. We can
estimate that the ratio between the SC pairing order parameter in
$(\mathbf{k},-\mathbf{k}+\mathbf{Q})$ channel and that in $(\mathbf{k}%
,-\mathbf{k})$ channel should be $\sim$ 0.025. Thus, the atomic spin-orbital
coupling plays a neglectable\ role in SC states, and the Ref.
\cite{Kreisel2013PRB} verified this point in A$_{x}$Fe$_{2-y}$Se$_{2}$. Other
issues, such as the coupling between the monolayer FeSe and substrate STO,
could also break the glide-plane symmetry. Such couplings are tunable and
strongly affected by the fabrication process and the substrate
materials\cite{Peng2014NC,Peng2014PRL}. Here, we consider the case that the
strength of coupling between the monolayer FeSe and substrate is weak in
comparison with the relevant hopping amplitude.

Compared with the general topological materials, in which the extended $s$ and
$p$ orbitals are the bricks to build low-energy electronic structures, and the
spin-orbital coupling plays an essential role in inducing the strong linear
couplings, the linear couplings in monolayer FeSe/STO is attributed to
effective couplings between 3d orbitals induced by d-p hybridizations from the
unique non-symmorphic lattice structures. Such features provide us an
alternative route to search for the new topological materials in
strongly-correlative electronic systems.

In conclusion, we propose that the monolayer FeSe/STO could support the
odd-parity topological SC states with the nodeless s-wave gap structures. In
contrast with other topological superconductors\cite{Fu2008PRL,Fu2010PRL}in
which the spin-orbital coupling plays a key role, such topological SC states
have strong relations with the unique non-symmorphic lattice symmetry which
induces the orbital-momentum locking. Furthermore, we calculate phase diagram
and suggest some experimental schemes to identify such uniquely nontrivial
topological SC states.

\begin{acknowledgments}
We thank Prof. J. P. Hu for helpful discussions. This work is supported by
Research Grants Council, University Grants Committee, Hong Kong, under HKU703713p.
\end{acknowledgments}

\appendix

\section{The tight-binding Hamiltonian from symmetry analyses}

In this section, we discuss the properties of the tight-binding Hamiltonian
from the symmetric point. The trilayer structure of the monolayer FeSe is
shown in Fig.1 (see main text). We focus on the three space group operations
including glide plane symmetry operator, $\hat{g}_{z}=\left\{  m_{z}%
|\mathbf{r}_{0}\right\}  $ with $\mathbf{r}_{0}=(\frac{1}{2}\frac{1}{2})$ and
two reflection symmetry operations, $\hat{g}_{x}=\left\{  m_{x}|\mathbf{r}%
_{0}\right\}  $ and $\hat{g}_{x^{\prime}}=\left\{  m_{x^{\prime}}|00\right\}
$. Besides, the lattice has inverse symmetry denoted by the operator, $\hat
{g}_{i}=\left\{  i|\mathbf{r}_{0}\right\}  $. According to the LDA
calculation, we can only focus on Fe atoms, the Bloch wave functions for the
3$d$ orbitals of Fe are defined as%
\begin{equation}
|\alpha\eta,\mathbf{k}^{\prime}\rangle=\frac{1}{\sqrt{N}}\sum_{n}%
e^{i\mathbf{k}^{\prime}\cdot\mathbf{r}_{n\eta}^{\prime}}\phi_{\alpha
}(\mathbf{r}^{\prime}-\mathbf{r}_{n\eta}^{\prime}). \label{basisfunc1}%
\end{equation}
Here, $\mathbf{r}_{n\eta}^{\prime}=\mathbf{R}_{n}^{\prime}+\mathbf{r}_{\eta
}^{\prime}$ \bigskip with lattice vector $\mathbf{R}_{n}^{\prime}$ and the
position $\mathbf{r}_{\eta}^{\prime}$ of Fe atom $\eta=A,B$, and $\phi
_{\alpha}$ denotes the d orbital basis function ($\alpha=xz,yz,x^{2}%
-y^{2},xy,z^{2}$). The symmetry operators acting on the basis function
$|\alpha\eta,\mathbf{k}^{\prime}\rangle$ have the following properties,%
\begin{align}
\hat{g}_{x^{\prime}}|\alpha\eta,\mathbf{k}^{\prime}\rangle &  =\sum_{\beta
}m_{x^{\prime},\alpha\beta}|\beta\eta,m_{x^{\prime}}\mathbf{k}^{\prime}%
\rangle\nonumber\\
\hat{g}_{z}|\alpha\eta,\mathbf{k}^{\prime}\rangle &  =\sum_{\beta}%
e^{-i(\hat{m}_{z}\mathbf{k}^{\prime})\cdot\mathbf{r}_{0}}m_{z,\alpha\beta
}|\beta\bar{\eta},\hat{m}_{z}\mathbf{k}^{\prime}\rangle\nonumber\\
\hat{g}_{x}|\alpha\eta,\mathbf{k}^{\prime}\rangle &  =\sum_{\beta}%
e^{-i(\hat{m}_{x}\mathbf{k}^{\prime})\cdot\mathbf{r}_{0}}m_{x,\alpha\beta
}|\beta\bar{\eta},\hat{m}_{x}\mathbf{k}^{\prime}\rangle. \label{Tans}%
\end{align}
The relevant tight-binding (TB) Hamiltonian can be expressed as%
\begin{equation}
H_{0}=\sum_{\mathbf{k}^{\prime}}\Psi^{\dag}(\mathbf{k}^{\prime})H(\mathbf{k}%
^{\prime})\Psi(\mathbf{k}^{\prime}), \label{Tb1}%
\end{equation}
with%
\begin{align}
\Psi^{\dag}(\mathbf{k}^{\prime})  &  =[\psi_{A}^{\dag}(\mathbf{k}^{\prime
}),\psi_{B}^{\dag}(\mathbf{k}^{\prime})]\label{Tbbasis}\\
\psi_{\eta}^{\dag}(\mathbf{k}^{\prime})  &  =[d_{\eta,xz}^{\dag}%
(\mathbf{k}^{\prime}),d_{\eta,yz}^{\dag}(\mathbf{k}^{\prime}),d_{\eta
,x^{2}-y^{2}}^{\dag}(\mathbf{k}^{\prime}),d_{\eta,xy}^{\dag}(\mathbf{k}%
^{\prime}),d_{\eta,z^{2}}^{\dag}(\mathbf{k}^{\prime})].\nonumber
\end{align}
In the basis $\Psi(\vec{k}^{\prime})$, the corresponding transformation
matrices for the three operations $\hat{g}_{\alpha}$ have the following forms,%
\begin{align}
U(\hat{g}_{x^{\prime}})  &  =\left[
\begin{array}
[c]{cc}%
m_{x^{\prime}} & 0\\
0 & m_{x^{\prime}}%
\end{array}
\right] \nonumber\\
U(\hat{g}_{z})  &  =\left[
\begin{array}
[c]{cc}%
0 & e^{-i(m_{z}\mathbf{k}^{\prime})\cdot\mathbf{r}_{0}}m_{z}\\
e^{-i(m_{z}\mathbf{k}^{\prime})\cdot\mathbf{r}_{0}}m_{z} & 0
\end{array}
\right] \nonumber\\
U(\hat{g}_{x})  &  =\left[
\begin{array}
[c]{cc}%
0 & e^{-i(m_{x}\mathbf{k}^{\prime})\cdot\mathbf{r}_{0}}m_{x}\\
e^{-i(m_{z}\mathbf{k}^{\prime})\cdot\mathbf{r}_{0}}m_{x} & 0
\end{array}
\right]  , \label{Trans2}%
\end{align}
Where%
\begin{align}
m_{x^{\prime}}  &  =\left[
\begin{array}
[c]{ccccc}%
0 & 1 & 0 & 0 & 0\\
1 & 0 & 0 & 0 & 0\\
0 & 0 & -1 & 0 & 0\\
0 & 0 & 0 & 1 & 0\\
0 & 0 & 0 & 0 & 1
\end{array}
\right]  ,m_{z}=\left[
\begin{array}
[c]{ccccc}%
-1 & 0 & 0 & 0 & 0\\
0 & -1 & 0 & 0 & 0\\
0 & 0 & 1 & 0 & 0\\
0 & 0 & 0 & 1 & 0\\
0 & 0 & 0 & 0 & 1
\end{array}
\right]  ,\label{Matrx1}\\
m_{x}  &  =\left[
\begin{array}
[c]{ccccc}%
-1 & 0 & 0 & 0 & 0\\
0 & 1 & 0 & 0 & 0\\
0 & 0 & 1 & 0 & 0\\
0 & 0 & 0 & -1 & 0\\
0 & 0 & 0 & 0 & 1
\end{array}
\right]  .\nonumber
\end{align}
The symmetry of the Hamiltonian requires%
\begin{equation}
H_{0}(\mathbf{k}^{\prime})=U(\mathbf{k}^{\prime})H_{0}(U\mathbf{k}^{\prime
})U^{\dag}(\mathbf{k}^{\prime}). \label{TB2}%
\end{equation}
Define%
\begin{equation}
H_{0}(\mathbf{k}^{\prime})=\left[
\begin{array}
[c]{cc}%
H_{A}(\mathbf{k}^{\prime}) & H_{AB}(\mathbf{k}^{\prime})\\
H_{BA}(\mathbf{k}^{\prime}) & H_{B}(\mathbf{k}^{\prime})
\end{array}
\right]  . \label{TB2cone}%
\end{equation}
We can get%
\begin{align}
H_{A/B}(k_{x^{\prime}},k_{y^{\prime}})  &  =m_{x^{\prime}}H_{A/B}%
(-k_{x^{\prime}},k_{y^{\prime}})m_{x^{\prime}}\nonumber\\
H_{AB}(k_{x^{\prime}},k_{y^{\prime}})  &  =m_{x^{\prime}}H_{AB}(-k_{x^{\prime
}},k_{y^{\prime}})m_{x^{\prime}} \label{trans_1}%
\end{align}%
\begin{align}
H_{A}(k_{x^{\prime}},k_{y^{\prime}})  &  =m_{z}H_{B}(k_{x^{\prime}%
},k_{y^{\prime}})m_{z}\nonumber\\
H_{AB}(k_{x^{\prime}},k_{y^{\prime}})  &  =m_{z}H_{BA}(k_{x^{\prime}%
},k_{y^{\prime}})m_{z} \label{trans_2}%
\end{align}%
\begin{align}
H_{A}(k_{x^{\prime}},k_{y^{\prime}})  &  =m_{x}H_{B}(-k_{y^{\prime}%
},-k_{x^{\prime}})m_{x}\nonumber\\
H_{AB}(k_{x^{\prime}},k_{y^{\prime}})  &  =m_{x}H_{BA}(-k_{y^{\prime}%
},-k_{x^{\prime}})m_{x}. \label{trans_3}%
\end{align}
Moreover, since $|\alpha\eta,\mathbf{k}^{\prime}+\mathbf{G}^{\prime}%
\rangle=e^{i\mathbf{G}^{\prime}\cdot\mathbf{r}_{\eta}^{\prime}}|\alpha
\eta,\mathbf{k}^{\prime}\rangle$%
\begin{align}
H_{A/B}(\mathbf{k}^{\prime}+\mathbf{G}^{\prime})  &  =H_{A/B}(\mathbf{k}%
^{\prime})\nonumber\\
H_{AB}(\mathbf{k}^{\prime}+\mathbf{G}^{\prime})  &  =e^{i\mathbf{G}^{\prime
}\cdot\mathbf{r}_{0}^{\prime}}H_{AB}(\mathbf{k}^{\prime}). \label{sy1}%
\end{align}
$\mathbf{r}_{0}^{\prime}=\mathbf{r}_{B}^{\prime}-\mathbf{r}_{A}^{\prime
}=(\frac{1}{2},\frac{1}{2})$ Considering the operator $\hat{g}_{z}$, we can
find in the entire BZ%
\begin{equation}
\left[  \left[
\begin{array}
[c]{cc}%
0 & m_{z}\\
m_{z} & 0
\end{array}
\right]  ,\left[
\begin{array}
[c]{cc}%
H_{A}(\mathbf{k}^{\prime}) & H_{AB}(\mathbf{k}^{\prime})\\
H_{BA}(\mathbf{k}^{\prime}) & H_{B}(\mathbf{k}^{\prime})
\end{array}
\right]  \right]  =0. \label{cal1}%
\end{equation}
We have%
\begin{equation}
V^{\dag}\left[
\begin{array}
[c]{cc}%
0 & m_{z}\\
m_{z} & 0
\end{array}
\right]  V=\left[
\begin{array}
[c]{cc}%
-I_{5\times5} & 0\\
0 & I_{5\times5}%
\end{array}
\right]  . \label{cal2}%
\end{equation}%
\begin{equation}
V=\frac{1}{\sqrt{2}}\left[
\begin{array}
[c]{cc}%
A & A\\
B & -B
\end{array}
\right]  , \label{cal3}%
\end{equation}
with $A=I_{5\times5},B=-m_{z}$. It is straightforward to check that
$H_{0}(\mathbf{k}^{\prime})$ can also be block diagonalized, i.e.,%

\begin{equation}
V^{\dag}H_{0}(\mathbf{k}^{\prime})V=H_{11}(\mathbf{k}^{\prime})\oplus
H_{22}(\mathbf{k}^{\prime}), \label{transend}%
\end{equation}
with $H_{11}(\mathbf{k}^{\prime})=H_{A}(\mathbf{k}^{\prime})-H_{AB}%
(\mathbf{k}^{\prime})m_{z}$ and $H_{22}(\mathbf{k}^{\prime})=H_{A}%
(\mathbf{k}^{\prime})+H_{AB}(\mathbf{k}^{\prime})m_{z}$. From Eq.(\ref{sy1}),
we can get $H_{A/B}(k_{x^{\prime}}+2\pi n_{x^{\prime}},k_{x^{\prime}}+2\pi
n_{y^{\prime}})=H_{A/B}(k_{x^{\prime}}+2\pi n_{x^{\prime}},k_{x^{\prime}}+2\pi
n_{y^{\prime}})$ and $H_{AB}(k_{x^{\prime}}+2\pi n_{x^{\prime}},k_{x^{\prime}%
}+2\pi n_{y^{\prime}})=e^{i(2\pi n_{x^{\prime}}\frac{1}{2}+2\pi n_{y^{\prime}%
}\frac{1}{2})}H_{AB}(k_{x^{\prime}}+2\pi n_{x^{\prime}},k_{x^{\prime}}+2\pi
n_{y^{\prime}})$. When $(n_{x^{\prime}},n_{y^{\prime}})=(0,1)$, $H_{11}%
(\mathbf{k}^{\prime})=H_{A}(\mathbf{k}^{\prime})-H_{AB}(\mathbf{k}^{\prime
})m_{z}$ and $H_{22}(\mathbf{k}^{\prime})=H_{A}(\mathbf{k}^{\prime}%
+\mathbf{Q}^{\prime})-H_{AB}(\mathbf{k}^{\prime}+\mathbf{Q}^{\prime})m_{z}$
with $\mathbf{Q}^{\prime}=(0,2\pi)$. Furthermore, the momentum defined in the
one-Fe BZ is $k_{x}=(k_{x^{\prime}}+k_{y^{\prime}})/2$, $k_{y}=(-k_{x^{\prime
}}+k_{y^{\prime}})/2$ and $\mathbf{Q}=(\pi,\pi).$

Under the basis, $\Psi^{\dag}(\mathbf{k})=[\psi^{\dag}(\mathbf{k}),\psi^{\dag
}(\mathbf{k}+\mathbf{Q})]$ with $\psi^{\dag}(\mathbf{k})=[d_{xz}^{\dag
}(\mathbf{k}),d_{yz}^{\dag}(\mathbf{k}),d_{x^{2}-y^{2}}^{\dag}(\mathbf{k}%
),d_{xy}^{\dag}(\mathbf{k}),d_{z^{2}}^{\dag}(\mathbf{k})]$, $d_{l}%
(\mathbf{k})=\frac{1}{\sqrt{2}}[d_{A,l}(\mathbf{k}^{\prime})+d_{B,l}%
(\mathbf{k}^{\prime})]$ and $d_{l}(\mathbf{k}+\mathbf{Q})=\frac{1}{\sqrt{2}%
}[d_{A,l}(\mathbf{k}^{\prime})-d_{B,l}(\mathbf{k}^{\prime})]$ for $l=xz,yz$,
$d_{l}(\mathbf{k})=\frac{1}{\sqrt{2}}[d_{A,l}(\mathbf{k}^{\prime}%
)-d_{B,l}(\mathbf{k}^{\prime})]$ and $d_{l}(\mathbf{k}+\mathbf{Q})=\frac
{1}{\sqrt{2}}[d_{A,l}(\mathbf{k}^{\prime})+d_{B,l}(\mathbf{k}^{\prime})]$ for
$l=xy,x^{2}-y^{2},z^{2}$, the TB Hamiltonian in the one-Fe BZ takes the
following form%
\begin{equation}
H_{0}=\sum_{\mathbf{k}}\Psi^{\dag}(\mathbf{k})H_{0}(\mathbf{k})\Psi
(\mathbf{k}). \label{TB1Fe}%
\end{equation}
Then,%

\begin{equation}
H_{0}(\mathbf{k})=H_{o}(\mathbf{k})\oplus H_{e}(\mathbf{k}). \label{Tb1Fe1}%
\end{equation}
Here, $H_{e}(\mathbf{k})=H_{o}(\mathbf{k}+\mathbf{Q})$.

The TB Hamiltonian in one-Fe BZ Eq. (\ref{Tb1Fe1}) have block-diagonal forms,
and each block has definitive orbital parity with respect to the glide plane
symmetry. Besides, the inversion symmetry $\hat{g}_{i}=\left\{  i|\mathbf{r}%
_{0}\right\}  $ indicates that the inversion center of monolayer FeSe is at
the midpoint of Fe-Fe link. Thus, we can find that $d_{xz/yz}(\mathbf{k}%
)$/$d_{xy/x^{2}-y^{2}/z^{2}}(\mathbf{k})$ are inversion even/odd, and
$d_{xz/yz}(\mathbf{k+Q})$/$d_{xy/x^{2}-y^{2}/z^{2}}(\mathbf{k+Q})$ are
inversion odd/even. In other words, $d_{xz/yz}$ orbitals and $d_{xy/x^{2}%
-y^{2}/z^{2}}$ orbitals have opposite parities in the subspace with definitive
orbital parity. The TB Hamiltonian in the one-Fe BZ is%
\begin{equation}
H_{o}(\vec{k})=\left[
\begin{array}
[c]{ccccc}%
A_{11} & A_{12} & A_{13} & A_{14} & A_{15}\\
& A_{22} & A_{23} & A_{24} & A_{25}\\
&  & A_{33} & A_{34} & A_{35}\\
&  &  & A_{44} & A_{45}\\
&  &  &  & A_{55}%
\end{array}
\right]  . \label{TBodd}%
\end{equation}
The non-zero terms in $A(k)$ are listed as follows,%

\begin{align*}
A_{11/22}(k)  &  =\epsilon_{1}+2t_{x/y}^{11}\cos k_{x}+2t_{y/x}^{11}\cos
k_{y}+4t_{xy}^{11}\cos k_{x}\cos k_{y}\\
&  +2t_{xx/yy}^{11}\cos2k_{x}+2t_{yy/xx}^{11}\cos2k_{y}\\
&  +4t_{xxy/yyx}^{11}\cos2k_{x}\cos k_{y}+4t_{xyy/xxy}^{11}\cos k_{x}%
\cos2k_{y}\\
&  +4t_{xxyy}^{11}\cos2k_{x}\cos2k_{y},
\end{align*}

\[
A_{33}(k)=\epsilon_{3}+2t_{x}^{33}(\cos k_{x}+\cos k_{y})+4t_{xy}^{33}\cos
k_{x}\cos k_{y},
\]

\begin{align*}
A_{44}(k)  &  =\epsilon_{4}+2t_{x}^{44}(\cos k_{x}+\cos k_{y})+4t_{xy}%
^{44}\cos k_{x}\cos k_{y}\\
&  +4t_{xxy}^{44}(\cos2k_{x}\cos k_{y}+\cos k_{x}\cos2k_{y})\\
&  +4t_{xxyy}^{44}\cos2k_{x}\cos2k_{y},
\end{align*}%
\[
A_{55}(k)=\epsilon_{5},
\]%
\[
A_{12}(k)=-4t_{xy}^{12}\sin k_{x}\sin k_{y},
\]%
\[
A_{13/23}(k)=\pm2it_{x}^{13}\sin k_{y/x}\pm4it_{xy}^{13}\sin k_{y/x}\cos
k_{x/y},
\]%
\[
A_{14/24}(k)=-2it_{x}^{14}\sin k_{x/y}+4it_{xy}^{14}\sin k_{x/y}\cos k_{y/x},
\]%
\[
A_{15/25}(k)=2it_{x}^{15}\sin k_{y/x}+4it_{xy}^{15}\sin k_{y/x}\cos k_{x/y},
\]%
\[
A_{35}(k)=2t_{x}^{35}(\cos k_{x}-\cos k_{y}),
\]%
\[
A_{45}(k)=-4t_{xy}^{45}\sin k_{x}\sin k_{y}.
\]

The onsite orbital energy is $\epsilon_{1}=\epsilon_{2}=0.02.$ $\epsilon
_{3}=-0.539.\epsilon_{4}=0.014.\epsilon_{5}=-0.581$, and the hopping
parameters for the free-standing monolayer FeSe are listed as
follows\cite{Eschrig}, $t_{x/y}^{11}=-0.08/-0.311$, $t_{xy}^{11}=0.232$,
$t_{xx/yy}^{11}=0.009/-0.045$, $t_{xxy/yyx}^{11}=-0.016/0.019$, $t_{xxyy}%
^{11}=0.02$, $t_{x}^{33}=0.412$, $t_{xy}^{33}=-0.066$, $t_{x}^{44}=0.063$,
$t_{xy}^{44}=0.086$, $t_{xxy}^{44}=-0.017$, $t_{xxyy}^{44}=-0.028$,
$t_{xy}^{12}=0.099$, $t_{x}^{13}=0.3$, $t_{xy}^{13}=-0.089$, $t_{x}%
^{14}=0.305$, $t_{xy}^{13}=-0.056$, $t_{x}^{15}=-0.18$, $t_{xy}^{15}=0.146$,
$t_{x}^{35}=0.338$, $t_{xy}^{45}=-0.109$. The renormalized parameters
corresponding to Fig. 1 (d) in main text are $t_{xy}^{44}=0.066$, $t_{x}%
^{14}=0.405$, $t_{x}^{11}=-0.12$. The renormalized parameters corresponding to
Fig. 1 (e) in main text are $t_{xy}^{44}=0.076$, $t_{x}^{44}=0.183$,
$t_{x}^{14}=0.405$, $t_{x}^{11}=-0.311$, $t_{xy}^{11}=0.19$.

\section{The classifications for the ($\mathbf{k},-\mathbf{k}+\mathbf{Q}$)
pairing channels from symmetry analyses}

The eight GellMann matrices $\lambda_{0}\sim\lambda_{8}$ in the main text are
listed as follows,%
\begin{align}
\lambda_{0}  &  =\left[
\begin{array}
[c]{ccc}%
1 & 0 & 0\\
0 & 1 & 0\\
0 & 0 & 1
\end{array}
\right]  ,\lambda_{1}=\left[
\begin{array}
[c]{ccc}%
0 & 1 & 0\\
1 & 0 & 0\\
0 & 0 & 0
\end{array}
\right]  ,\lambda_{2}=\left[
\begin{array}
[c]{ccc}%
0 & -i & 0\\
i & 0 & 0\\
0 & 0 & 0
\end{array}
\right] \nonumber\\
\lambda_{3}  &  =\left[
\begin{array}
[c]{ccc}%
1 & 0 & 0\\
0 & -1 & 0\\
0 & 0 & 0
\end{array}
\right]  ,\lambda_{4}=\left[
\begin{array}
[c]{ccc}%
0 & 0 & 1\\
0 & 0 & 0\\
1 & 0 & 0
\end{array}
\right]  ,\lambda_{5}=\left[
\begin{array}
[c]{ccc}%
0 & 0 & -i\\
0 & 0 & 0\\
i & 0 & 0
\end{array}
\right] \nonumber\\
\lambda_{6}  &  =\left[
\begin{array}
[c]{ccc}%
0 & 0 & 0\\
0 & 0 & 1\\
0 & 1 & 0
\end{array}
\right]  ,\lambda_{7}=\left[
\begin{array}
[c]{ccc}%
0 & 0 & 0\\
0 & 0 & -i\\
0 & i & 0
\end{array}
\right]  ,\lambda_{8}=\frac{1}{\sqrt{3}}\left[
\begin{array}
[c]{ccc}%
1 & 0 & 0\\
0 & 1 & 0\\
0 & 0 & -2
\end{array}
\right]  . \label{Gell}%
\end{align}

\begin{table}[ptb]
\caption{The IRs of all the possible onsite superconducting pairing in
$(\mathbf{k},-\mathbf{k}+\mathbf{Q})$channels}%
\begin{tabular}
[c]{llllll}\hline\hline
$(\mathbf{k},-\mathbf{k}+\mathbf{Q}):\Delta^{\prime}(\mathbf{k})$ & $c_{2}(z)$
& $c_{2}(x)$ & $\sigma_{d}$ & $\{i|\frac{1}{2}\frac{1}{2}\}^{\prime}$ &
$IR$\\\hline
$s_{0}\lambda_{0}$ & 1 & 1 & 1 & -1 & $A_{1u}$\\
$s_{0}\lambda_{8}$ & 1 & 1 & 1 & -1 & $A_{1u}$\\
$s_{0}\lambda_{1}$ & 1 & -1 & 1 & -1 & $B_{2u}$\\
$s_{0}(\lambda_{4},$ $\lambda_{6})$ & (-1,-1) & (1,-1) & $s_{0}(\lambda_{6},$
$\lambda_{4})$ & (1,1) & $E_{g}$\\
$is_{z}\lambda_{2}$ & 1 & 1 & 1 & -1 & $A_{1u}$\\
$s_{z}(\lambda_{5},\lambda_{7})$ & (-1,-1) & (-1,1) & -$s_{z}(\lambda
_{7},\lambda_{5})$ & (1,1) & $E_{g}$\\
$i(s_{x},s_{y})\lambda_{2}$ & (-1,-1) & (-1,1) & $i(s_{y},s_{x})\lambda_{2}$ &
(-1,-1) & $E_{u}$\\
$i(s_{x}\lambda_{5},s_{y}\lambda_{7})$ & (1,1) & (1,1) & -$i(s_{y}\lambda
_{7},s_{x}\lambda_{5})$ & (1,1) & $E_{g}$\\
$i(s_{y}\lambda_{5},s_{x}\lambda_{7})$ & (1,1) & (-1,-1) & -$i(s_{x}%
\lambda_{7},s_{y}\lambda_{5})$ & (1,1) & $E_{g}$\\\hline\hline
\end{tabular}
\end{table}

\begin{table}[ptb]
\caption{The IRs of all the possible non-onsite superconducting pairing in
$(\mathbf{k},-\mathbf{k}+\mathbf{Q})$ channels }%
\begin{tabular}
[c]{ll}\hline\hline
$(\mathbf{k},-\mathbf{k}+\mathbf{Q}):\Delta^{\prime}(\mathbf{k})$ & IR\\\hline
$f_{4,k}s_{0}\lambda_{0/8},f_{5,k}s_{0}\lambda_{1},f_{3,k_{x}}s_{0}\lambda
_{5}+f_{3,k_{y}}s_{0}\lambda_{7}$ & $A_{1u}$\\
$f_{2,k}s_{0}\lambda_{0/8},f_{3,k_{x}}s_{0}\lambda_{5}-f_{3,k_{y}}s_{0}%
\lambda_{7}$ & $B_{1u}$\\
$f_{2,k}s_{0}\lambda_{1},f_{3,k_{y}}s_{0}\lambda_{5}-f_{3,k_{x}}s_{0}%
\lambda_{7}$ & $A_{2u}$\\
$f_{5,k}s_{0}\lambda_{0/8},f_{1/4,k}s_{0}\lambda_{1},f_{3,k_{y}}s_{0}%
\lambda_{5}+f_{3,k_{x}}s_{0}\lambda_{7}$ & $B_{2u}$\\\hline
$if_{1/4,k}s_{z}\lambda_{2},i^{1/0/0}[f_{3,k_{x}}s_{z/x/y}\lambda
_{4}+f_{3,k_{y}}s_{z/y/x}\lambda_{6}]$ & $A_{1u}$\\
$if_{2,k}s_{z}\lambda_{2},i^{1/0/0}[f_{3,k_{x}}s_{z/x/y}\lambda_{4}%
-f_{3,k_{y}}s_{z/y/x}\lambda_{6}]$ & $B_{1u}$\\
$i^{1/0/0}[f_{3,k_{y}}s_{z/x/y}\lambda_{4}-f_{3,k_{x}}s_{z/y/x}\lambda_{6}]$ &
$A_{2u}$\\
$if_{5,k}s_{z}\lambda_{2},i^{1/0/0}[f_{3,k_{y}}s_{z/x/y}\lambda_{4}%
+f_{3,k_{x}}s_{z/y/x}\lambda_{6}]$ & $B_{2u}$\\
$if_{1/2/4/5,k}(s_{x},s_{y})\lambda_{2}$ & $E_{u}$\\\hline
\end{tabular}
\end{table}

The monolayer FeSe has inversion symmetry, thus every IR in Table I of main
text should have a counterpart with a inverse parity. In other words,
$(\mathbf{k},-\mathbf{k}+\mathbf{Q})$ pairing channels should be possible from
the symmetry point. For the $(\mathbf{k},-\mathbf{k}+\mathbf{Q})$ pairing, We
define the the Nambu basis, $\Psi^{\prime}(\mathbf{k})=[\{\psi_{m\uparrow
}(\mathbf{k})\},\{\psi_{m\downarrow}(\mathbf{k})\},\{\psi_{m\downarrow}^{\dag
}(-\mathbf{k+Q})\},-\{\psi_{m\uparrow}^{\dag}(-\mathbf{k+Q})\}]^{t}$, with
$\{\psi_{m\sigma}(\mathbf{k})\}=[d_{xz\sigma}(\mathbf{k}),d_{yz\sigma
}(\mathbf{k}),d_{xy\sigma}(\mathbf{k})]$. The IRs for the onsite
$(\mathbf{k},-\mathbf{k}+\mathbf{Q})$ pairings are summarized in Table III.
Here the matrix for $\{i|\frac{1}{2}\frac{1}{2}\}^{\prime}$ is $g_{4}^{\prime
}=s_{0}\eta_{4}^{\prime}$ and $\eta_{4}^{\prime}=1\oplus-1\oplus(-1)^{\alpha}$
with $\alpha=1$ for d$_{xz}$-d$_{xy}$ pairing and $\alpha=-1$ for d$_{yz}%
$-d$_{xy}$ pairing. The IRs for the non-onsite $(\mathbf{k},-\mathbf{k}%
+\mathbf{Q})$ pairings are summarized in Table IV. We can check that all the
$(\mathbf{k},-\mathbf{k}+\mathbf{Q})$ pairing channels correspond to the
inter-band pairings, and such kinds of pairings can not individually give a
overall full gap\ around the Fermi surface.


\begin{thebibliography}{99}                                                                                               %


\bibitem {Qi2011RMP}X.-L. Qi and S.-C. Zhang, Rev. Mod. Phys. \textbf{83,}
1057--1110 (2011).

\bibitem {Fu2008PRL}L. Fu and C. L. Kane, Phys. \ Rev. Lett. \textbf{100,}
096407 (2008).

\bibitem {Sau2010PRL}J. D. Sau, R. M Lutchyn, S. Tewari and S. Das Sarma,
Phys. Rev. Lett. \textbf{104,} 040502 (2010).

\bibitem {Fu2010PRL}L. Fu and E. Berg, Phys. Rev. Lett. \textbf{105,} 097001 (2010).

\bibitem {Yoichi2008JACS}Y. Kamihara, T. Watanabe, M. Hirano and H.
Hosono\textit{,} J. Am. Chem. Soc. \textbf{130,} 3296-3297 (2008).

\bibitem {Yin2015NP}J.-X Yin, Z. Wu, J.-H. Wang, Z.-Y. Ye, J. Gong, X. -Y.
Hou, L. Shan, A. Li, X.-J. Liang, X.-X. Wu, J. Li, C.-S. Ting, Z. Wang, J.-P.
Hu, P.-H. Hor, H. Ding and S. H. Pan, Nat. Phys. \textbf{11,} 543 (2015).

\bibitem {Hao2014PRX}N. Hao and J. Hu, Phys. Rev. X \textbf{4,} 031053 (2014).

\bibitem {Wu2014arXiv}X. Wu, S. Qin, Y. Liang, H. Fan, and J. Hu,
arXiv:1412.3375 (2014).

\bibitem {Wang2012CPL}Q.-Y. Wang \textit{et al.}, Chinese Phys. Lett.
\textbf{29,} 037402 (2012).

\bibitem {Liu2012NC}D. Liu \textit{et al.},\textit{ }Nat. Commun. \textbf{3,}
931 (2012).

\bibitem {He2013NM}S. He \textit{et al.,} Nat. Mater. \textbf{12,} 605--610 (2013).

\bibitem {Tan2013NM}S. Tan \textit{et al.}, Nat. Mater. \textbf{12,} 634--640 (2013).

\bibitem {Peng2014NC}R. Peng \textit{et al.}, Nat. Commun. \textbf{5, }5044 (2014).

\bibitem {Zhang2014CPL}W.-H. Zhang \textit{et al.},\textit{ }Chinese Phys.
Lett. \textbf{31,} 017401 (2014).

\bibitem {Lee2014N}J. J. Lee \textit{et al.}, Nature \textbf{515,} 245--248 (2014).

\bibitem {Ge2015NM}J.-F. Ge \textit{et al.}, Nat. Mater\textit{.} \textbf{14},
285--289 (2015).

\bibitem {Hor2010PRL}Y. S. Hor \textit{et al.}, Phys. Rev. Lett.\textit{
}\textbf{104,} 057001 (2010).

\bibitem {Sasaki2011PRL}S. Sasaki \textit{et al.}, Phys. Rev. Lett\textit{.}
\textbf{107,} 217001 (2011).

\bibitem {Sasaki2012PRL}S. Sasaki \textit{et al}, Phys. Rev. Lett\textit{.}
\textbf{109,} 217004 (2012).

\bibitem {Hu2013PRX}J. Hu, Phys. Rev. X \textbf{3,} 031004 (2013).

\bibitem {Hao2014PRB}N. Hao and J. Hu, Phys. Rev. B \textbf{89,} 045144 (2014).

\bibitem {Subedi2008PRB}A. Subedi, L. Zhang, D. J. Singh and M. H. Du, Phys.
Rev. B \textbf{78,} 134514 (2008).

\bibitem {Cvetkovic}V. Cvetkovic and O. Vafek, Phys. Rev. B \textbf{88},
134510 (2013).

\bibitem {Liu2012PRB}K. Liu, Z.-Y. Lu and T. Xiang, Phys. Rev. B \textbf{85,}
235123 (2012).

\bibitem {Timur2013JPCM}T. Bazhirov and M. L. Cohen, Journal of Physics:
Condensed Matter \textbf{25,} 105506 (2013).

\bibitem {Zheng2013SR}F. Zheng, Z. Wang, W. Kang and P. Zhang, Sci.
Rep\textit{.} \textbf{3, }2213 (2013).

\bibitem {Cao2008PRB}C. Cao, P. J. Hirschfeld and H.-P. Cheng, Phys. Rev. B
\textbf{77,} 220506 (2008).

\bibitem {Thouless1982PRL}D. J. Thouless, M. Kohmoto, M. P. Nightingale and M.
den Nijs, Phys. Rev. Lett. \textbf{49,} 405 (1982).

\bibitem {Sheng2006PRL}D. N. Sheng, Z. Y. Weng, L. Sheng and F. D. M. Haldane,
Phys. Rev. Lett\textit{.} \textbf{97,} 036808 (2006).

\bibitem {Wang2009PRB}Z. Wang, N. Hao and P. Zhang, Phys. Rev. B \textbf{80,}
115420 (2009).

\bibitem {Kane2005PRL}C. L. Kane and E. J. Mele\textit{,} Phys. Rev. Lett.
\textbf{95,} 146802 (2005).

\bibitem {Read2000PRB}N. Read and D. Green\textit{, }Phys. Rev. B \textbf{6}1,
10267 (2000).

\bibitem {Qi2009PRL}X.-L. Qi, T. L. Hughes, S. Raghu, and S.-C.
Zhang\textit{,} Phys. Rev. Lett. \textbf{102}, 187001 (2009).

\bibitem {Fu2012PRL-1}T. H. Hsieh and L. Fu, Phys. Rev. Lett. \textbf{108,}
107005 (2012).

\bibitem {Bozovic2014NP}I. Bozovic and C. Ahn\textit{,} Nat. Phys.
\textbf{10}, 892--895 (2014).

\bibitem {Maier2011PRB}T. A. Maier, S. Graser, P. J. Hirschfeld, and D. J.
Scalapino, Phys. Rev. B \textbf{83}, 100515(R) (2011).

\bibitem {Kreisel2013PRB}A. Kreisel, Y. Wang, T. A. Maier, P. J. Hirschfeld,
D. J. Scalapino, Phys. Rev. B. \textbf{88,} 094522 (2013).

\bibitem {ZhangNM2011}Y. Zhang, \textit{et al.}, Nat. Mater. \textbf{10}, 273
277 (2011).

\bibitem {XuPRB2012}M. Xu \textit{et al.}, Phys. Rev. B \textbf{85}, 220504 (2012).

\bibitem {Fang2011PRX}C. Fang, Y.-L. Wu, R. Thomale, B. A. Bernevig, and J.
Hu, Phys. Rev. \textbf{X} 1, 011009 (2011).

\bibitem {Park2011PRL}J. T. Park, G. Friemel, Yuan Li, J.-H. Kim, V. Tsurkan,
J. Deisenhofer, H.-A. Krug von Nidda, A. Loidl, A. Ivanov, B. Keimer, and D.
S. Inosov, Phys. Rev. Lett. \textbf{107}, 177005 (2011).

\bibitem {Bozovic2002PRL}I. Bozovic, G. Logvenov, I. Belca, B. Narimbetov and
I. Sveklo\textit{,} Phys. Rev. Lett. \textbf{89,} 107001 (2002).

\bibitem {Reyren2007S}N. Reyren \textit{et al.}, Science \textbf{317,}
1196--1199 (2007).

\bibitem {Ginzburg1964PL}V. L. Ginzburg, Phys. Lett. \textbf{13}, 101--102 (1964).

\bibitem {Rademaker2015Arxiv}L. Rademaker, Y. Wang, T. Berlijn, and S.
Johnston, arXiv:1507.03967 (2015).

\bibitem {Yi2015Arxiv}M. Yi \textit{et al.}, Arxiv:1505.06636 (2015).

\bibitem {Lee2008PRB}P. A. Lee and X.-G. Wen, Phys. Rev. B, \textbf{78} 144517 (2008).

\bibitem {Puetter2012EPL}C. M. Puetter and H.-Y. Kee, EPL, \textbf{98} 27010 (2012).

\bibitem {Anderson1987S}P. W. Anderson, Science \textbf{235}, 1196--1199 (1987).

\bibitem {Liu2015PRB}K. Liu, B.-J. Zhang and Z.-Y. Lu, Phys. Rev. B,
\textbf{91} 045107 (2015).

\bibitem {Hu2012SR}J. Hu and H. Ding, Sci. Rep. \textbf{2, }381 (2012).

\bibitem {Lu2014NP}Y.-M. Lu, T. Xiang and D.-H. Lee, Nat. Phys. \textbf{10},
634--637 (2014).

\bibitem {Hebel1959PR}L. C. Hebel and C. P. Slichter, Phys. Rev. \textbf{113},
1504 (1959).

\bibitem {Nagai2015Arxiv}Y. Nagai, Y. Ota, and M. Machida, arXiv:1504.08095 (2015).

\bibitem {Wollman1993PRL}D. A. Wollman, D. J. Van Harlingen, W. C. Lee, D. M.
Ginsberg and A. J. Leggett, Phys. Rev. Lett. \textbf{71}, 2134 (1993).

\bibitem {Hu2012PRX}J. Hu and N. Hao, Phys. Rev. X \textbf{2}, 021009 (2012).

\bibitem {Shiozaki2013PRL}K. Shiozaki and S. Fujimoto, Phys. Rev.
Lett\textit{.} \textbf{110,} 076804 (2013).

\bibitem {Yosuke2014arXiv}Y. Shimizu and K. Nomura,\textit{ }arXiv:1403.1021 (2014).

\bibitem {Peng2014PRL}R. Peng \textit{et al}, Phys. Rev. Lett. \textbf{112},
107001 (2014).

\bibitem {Eschrig}H. Eschrig and K. Koepernik, Phys. Rev. B \textbf{80},
104503 (2009).
\end{thebibliography}
\end{document}